\documentclass[amsmath, aps]{revtex4}
\usepackage{graphicx}
\usepackage{times}
\usepackage{epsfig}
\usepackage{amssymb}

\usepackage{amsmath}
\begin{document}
\title{Two-photon approximation in the theory of the electron recombination in hydrogen.}

\author{D. Solovyev$^1$ and L. Labzowsky$^{1,2}$}

\affiliation{ 
$^1$ V. A. Fock Institute of Physics, St. Petersburg
State University, Petrodvorets, Oulianovskaya 1, 198504,
St. Petersburg, Russia
\\
$^2$  Petersburg Nuclear Physics Institute, 188300, Gatchina, St.
Petersburg, Russia}
\begin{abstract}
A rigorous QED theory of the multiphoton decay of excited states in hydrogen atom is presented. The "two-photon" approximation is formulated which is limited by the one-photon and two-photon transitions including cascades transitions with two-photon links. This may be helpful for the strict description of the recombination process in hydrogen atom and, in principle, for the history of the hydrogen recombination in the early Universe.
\end{abstract}
\maketitle

\section{Introduction}

The recent accurate astrophysical observations and measurements of the cosmic microwave background (CMB) tempreture and polarization anisotropy \cite{Hin}, \cite{Page} triggered a new interest to the theory of the two-photon processes in hydrogen in view of the important role of these processes in the cosmological hydrogen recombination. The history of the hydrogen recombination in the early Universe is described in many reviews, for example \cite{Seager}. The bound-bound one-photon transitions from the upper levels to the lower ones did not permit the atoms to reach their ground states: each photon released in such a transition in one atom was immidiately absorbed by another one. In particular, the Lyman-alpha radiation 2p-1s, being reabsorbed, reemitted and again reabsorbed, did not allow the radiation to escape the interaction with the matter. As it was first established in \cite{Zeld}, \cite{Peebles} the two-photon 2s-1s radiative transition presents one of the main channels for the radiation escape from the interaction with matter. Hence, the recent properties of the CMB are essentially defined by the two-photon decay processes during the cosmological recombination epoch.

In \cite{Dubrovich}, \cite{Wong} it was argued that the $ns\rightarrow 1s$ ($n>2$) and $nd\rightarrow 1s$ two-photon transitions can also give a sizeable contribution to the process of the radiation escape from the interaction with the matter. Recently this problem was investigated thoroughly in the theoretical astrophysical studies in \cite{J.Chluba}, \cite{Hirata}. There is a crucial difference between the decay of the $ns$ ($n>2$) or $nd$ levels and the $2s$ decay level. This difference is due to the presence of the cascade transitions as the dominant decay channels in case of $ns$ ($n>2$) and $nd$ levels. For the $2s$ level the cascade transitions are absent. Since the cascade photons can be effectively reabsorbed, the problem of separation of the "pure" two-photon contribution from the cascade contribution arises. An interference between the two decay channels also should be taken into account. This problem appears to be not at all trivial and requires an application of the methods of the Quantum Electrodynamocs (QED) for the bound electrons.

Quantum Mechanical theory for the two-photon transitions was first developed by G\"{o}ppert-Mayer \cite{Goepp} and the first evaluation of the two-photon $2s\rightarrow 1s+2\gamma(E1)$ decay rate in hydrogen was performed by Breit and Teller \cite{Breit}. The accurate nonrelativistic calculation for this transition rate was given in \cite{Klarsfeld}; fully relativistic calculations, valid also for the H-like Highly Charged Ions (HCI) with arbitrary $Z$ (nuclear charge) values were performed in \cite{GD82}-\cite{Santos}. The most accurate recent calculation for this transition rate with the QED radiative corrections taken into account belongs to Jentschura \cite{Jent}. As well as for the neutral hydrogen, the cascade problem does not arise for the transition $2s\rightarrow 1s+2\gamma(E1)$ in the HCI with arbitrary $Z$ values.

The two-photon transitions were investigated theoretically and experimentally also in the few-electron and many-electron atoms and ions. In particular, the two-photon transition $1s2s\,^1S_0\rightarrow (1s)^2\,^1S_0+2\gamma(E1)$ transition rate for the neutral He atom was first evaluated in \cite{Dalg}. This decay channel also does not contain cascade contribution.

The cascade problem first did arise in connection with the decay of the metastable $2^3P_0$ level in He-like Uranium: $2^3P_0\rightarrow1^1S_0+\gamma(E1)+\gamma(M1)$. In this case there are two possible cascade transitions: $2^3P_0\rightarrow 2^3S_1+\gamma(E1)\rightarrow 1^1S_0+\gamma(E1)+\gamma(M1)$ and $2^3P_0\rightarrow 2^3P_1+\gamma(M1)\rightarrow 1^1S_0+\gamma(M1)+\gamma(E1)$. The corresponding decay rate was first evaluated by Drake \cite{Drake}. Later Savukov and Johnson \cite{Savukov} performed similar calculation for a variety of He-like ions ($50\leq Z\leq 92$). In \cite{Drake}, \cite{Savukov} the "pure" two-photon contribution was obtained by subtraction of a Lorentzian fir for the cascade contribution from the total two-photon decay frequency distribution. A rigorous QED approach for the evaluation of the two-photon decay probability in presence of cascades was developed in \cite{LabShon} on the basis of the Line Profile  Approach (LPA) in QED, i.e. the QED theory of the spectral line profile (see \cite{AndrLab}). The LPA consists of a standard evaluation of the decay probability as a transition probability to the lower levels. In the presence of cascades the integral over emitted photon frequency distribution becomes divergent due to the singular terms, corresponding to the cascade resonances. To avoid such a singularity, the resummation of an infinite series of the electron self-energy insertions into the electron propagator was performed in \cite{LabShon}. This resummation converts into a geometric progression and in this way the electron self-energy matrix element (and the level width as its imaginary part) enters the energy denominator and shifts the pole from the real axis into the complex energy plane, thus making the transition probability integral finite. With this approach F. Low \cite{Low} first derived the Lorentz profile from QED. In \cite{Drake}, \cite{Savukov} the level widths in the energy denominators were also introduced, though as the empirical parameters. In \cite{LabShon} the ambiguity of the separation of the "pure" two-photon decay and cascades was first revealed for HCI; it was shown also that the interference terms can essentially contribute to the total decay probability.

Nearly at the same time when the paper \cite{Drake} did arrive, the cascade problem was discussed also for the $ns$ ($n>2$), $nd$ transitions in the hydrogen atom \cite{cea86}, \cite{fsm88}. In these works the "pure" two-photon contribution was obtained simply by omitting the resonant  (singular) terms, responsible for the cascades. This approach was criticized later in \cite{J.Chluba}.  Another ("alternative") method which formally allows for the separate determination of the "pure" two-photon contribution in case of the two-photon transitions with cascades was developed in a series of works by U. Jentschura \cite{Jent1}-\cite{Jent3}. This approach contradicts to the LPA results. The LPA was applied to the $3s-1s$ transition (including cascade) in hydrogen in \cite{LSP}, where the ambiguity of the separation of the "pure" two-photon and the cascade contributions was again demonstrated numerically, as in the case of the HCI \cite{LabShon}. Very recently a paper \cite{Amaro} did arrive where an attempt was made to find a compromise between LPA and "alternative" approach. A reasonable agreement between the numerical results obtained by both methods was found. However, to our mind, the disagreement between the LPA and "alternative" approach is of conceptual character and cannot be eliminated.

Thus from the QED point of view only the total two-photon frequency distribution has a direct physical sense in case of the two-photon decays with cascades. This quantity should be a basic tool for the description of the two-photon processes in astrophysics. The employment of the "1+1" approximation for the description of cascades should be avoided. Along this way the most recent astrophysical theories \cite{J.Chluba}, \cite{Hirata} are built. Still the "1+1" approximation is not fully excluded from the considerations in \cite{J.Chluba}, \cite{Hirata}.

In view of the recent very accurate (with relative accuracy $\sim 1\%$) measurements of the properties of CMB \cite{Hin}, \cite{Page} and with expectation of the even more accurate ($\sim 0.1\%$) measurements in the near future, the theory of the cosmological recombination free of any uncertainties connected with the separation of the "pure" two-photon and cascade contributions should be formulated.

In the present paper we will formulate such a theory for the two-photon and the multiphoton decays in hydrogen. In this theory only two types of the level decays should be present: the direct one-photon decays when they are allowed and the total two-photon decays without separation of the "pure" two-photon decays and cascades. The total solution of the problem formulated above consists of two steps. First, the pure QED problem of the description of the multiphoton transitions in hydrogen in the "two-photon" approximation should be resolved. That is, all the decays of the excited levels should be classified and described either as the direct one-photon transitions to the ground state or as the two-photon transitions with cascades. In the "two-photon" approximation transitions with more than two nonresonant photons should be neglected. The formulation of the "two-photon" approximation should finalize the first step of the studies. The present paper will concern only this first step.

An important feature of the rigorous QED treatment of the process of recombination is that we have to trace the decay of every particular level up to the ground state. This is of course not the full picture of the recombination process. To be more close to the cosmological recombination one has to consider the transitions from the continuous states (plasma electrons) down to the ground state, taking into account the rescattering processes. This would correspond to the second step mentioned above. However, the accurate treatment of the recombination process from the particular excited level, as presented in this paper, also may be of interest. In particular, we demonstrate that the consequent QED treatment of the $3p$ level decay should include the two-photon contribution comparable with the widely discussed two-photon decay of $3s$ level \cite{cea86}-\cite{Amaro}. In this paper we limit ourselves only with electric dipole transitions (both in one- and two-photon decays) and ignore $n'd\rightarrow ns$ transitions which also are of importance (\cite{Dubrovich}-\cite{Hirata}).

At the second step one should modify the basic astrophysical equations describing the level population in hydrogen in such a way that the imput data for them should be, apart from the direct one-photon transition probabilities, only the total two-photon decay rates, including cascades, without separating out the "pure" two-photon decay rates. The use of the "1+1" approximation should be fully avoided. This task is beyond the scope of our paper.

Our paper is organized as follows. In Section II we formulate the basic concepts for the LPA-based theory for the two-photon decay with cascades. The two-photon approximation for the description of the multiphoton transitions in hydrogen is introduced. In Section III the standard derivation of the transition rate for the Lyman-$\alpha$ $2p-1s$ transition is presented and in Section IV the standard QED derivation of the Lorentz profile for this emission process is given. The same is done in Section V for the two-photon decay $2s-1s$. The decay of the $3s$ level is considered in Section VI and the ambiguity of separation of the "pure" two-photon and cascade contributions is demonstrated. The decay of $3p$ level in the two-photon approximation is described in Section VII, where it is shown that this decay also contains the two-photon contribution comparable with the two-photon contribution to the decay of $3s$-level. In Section VIII an investigation of the decay of $4s$ level in the "two-photon" approximation is performed which gives the clue to the general formulation of the two-photon approximation in the theory of the multiphoton transitions. Section IX contains discussion of the results and conclusions.

\section{Two-photon approximation for the multiphoton decays with cascades.}

In this Section we will follow the derivation in \cite{LSP} using this example for the formulation of the general principles of the "two-photon approximation" in the QED theory of the level decays. The grounds of this theory consist of few basic principles. First, all the decays should be traced up to the ground (stable) state. Within the "two-photon approximation" only such transitions can be defined unambigously. The two-photon approximation assumes that we take into account either direct (allowed) one-photon transitions from the excited level to the ground state, or the total two-photon transitions which end up also at the ground level. The two-photon transition rates consist of the inseparably mixed "pure" two-photon transitions and cascades. The contribution of cascades is dominating and can be of the order of the direct one-photon transitions. The "pure" two-photon contributions and the corresponding interference terms between the "pure" two-photon contributions and cascades define the level of accuracy of our theory: two-photon approximation. Thus, we neglect the "pure" three-photon, "pure" four-photon etc contributions, taking into account, however, the cascade parts of the multiphoton decays.

As the direct one-photon transitions as well as the cascade transitions we will consider only the allowed E1 transition dominant in the nonrelativistic theory. The order of magnitude of the corresponding transition rates parametrically equals to (in relativistic units) $W_{E1}^{(1\gamma)}=C_{E1}^{(1\gamma)}m\alpha (\alpha Z)^4$, where $m$ is the electron mass, $\alpha$ is the fine structure constant, $Z$ is the charge of nucleus and $C_{E1}^{(1\gamma)}$ is the numerical coefficient. In particular, for the Lyman-alpha $2p-1s$ transition in H-like ions $C_{E1}^{(\gamma 1)}(2p-1s)=732.722$.

Having in mind the astrophysical applications of our "two-photon" approximation we have to classify different decay channels with respect to their contribution to the radiation escape from the interaction with the matter. One of the main channels for this escape was already  mentioned above: this is the two-photon decay of the $2s$ state. The order of magnitude of the "pure" two-photon decay rate (in case of $2s$-level, when the cascade contribution is absent, the "pure" two-photon decay rate coincides with the total one) is $W^{(2\gamma)}_{E1E1}=C^{(2\gamma)}_{E1E1}m\alpha^2(\alpha Z)^6$ \cite{LKD}. In case of $2s\rightarrow 1s+2\gamma$ transition $C^{(2\gamma)}_{E1E1}(2s-1s)=24.7547$. We should stress that in the present paper we do not investigate in detail the process of the occupation of the $2s$ state. The situations when metastable state appears as an intermediate state in the cascade processes of transitions from the upper levels, will be included in the treatment of the cascades. In particular, in our treatment in this paper the total decay rate of the $3p$ level, apart from the one-photon decay rate $W^{(1\gamma)}_{E1}(3p-1s)$, incorporates also the two-photon decay rate $W_{E1E1}^{(2\gamma)}(2s-1s)$. This happens due to the existence of the cascade transition $3p\rightarrow 2s+\gamma\rightarrow 1s+3\gamma$. Then $\Gamma_{tot}(3p)=W_{E1}^{(1\gamma)}+\rm{const}\cdot W_{E1E1}^{(2\gamma)}$ (*) (see Section VI for details). In the standard treatment the transition rate of the upper  link of this cascade $W^{(1\gamma)}_{E1}(3p-2s)$ contributes directly to the $\Gamma_{tot}(3p)$: $\Gamma_{tot}(3p)=W_{E1}^{(1\gamma)}+W_{E1}^{(1\gamma)}(3p-2s)$ (**). The difference between these two situations can be explained by the two different types of the experiment. The equality (**) describes the laboratory experiment, when the photon with the frequency $\omega (3p-2s)$ is registered. Another situation occurs in the astrophysical context when it is important how fast an atom in an excited state will reach the ground level, i.e. the recombination will be accomplished. For a given cascade this depends on the slowest link of the cascade. In particular, for the $3p$ level the slowest link of the cascade $3p\rightarrow2s+1\gamma\rightarrow1s+3\gamma$ is the lower link: $2s\rightarrow1s+2\gamma$ and the total width $\Gamma_{tot}(3p)$ is defined by the equality (*). In the other words, one has to distinguish between the lifetime of a certain level $A$ (equation (**)) and the lifetime of an excited state of an atom, provided that initially this atom was in the state $A$ (equation (*)).

We remind that we consider here the decay processes in one single atom and ignore the possibility of reabsortion of the photon with the frequency $\omega=E(3p)-E(2s)$ by another atom.

\section{Decay rate for the 2p level in hydrogen.}

In Fig. 1 the decay scheme for $2p$ level (Lyman-alpha transition) is depicted. The emission process in frames of QED is described by the Feynman graph Fig. 2. The corresponding matrix element of the $S$-matrix is given by (see, for example \cite{LKD})
\begin{eqnarray}
\langle A'|\hat{S}^{(1)}|A\rangle = e \int d^4 x\, \bar{\psi}_{A'}(x)\gamma_{\mu}A^*_{\mu}(x)\psi_A(x)\,.
\end{eqnarray}
Here $\hat{S}^{(1)}$ is the first-order $S$-matrix, $e$ is the electron charge, $\psi_A(x) = \psi_A(\vec{r})e^{-i E_A t}$, $\psi_A(\vec{r})$ is the solution of the Dirac equation for the atomic electron, $E_A$ is the Dirac energy, $\bar{\psi}_{A'} = \psi_{A'}^\dagger \gamma_0$ is the Dirac conjugated wave function with $\psi_{A'}^{\dagger}$ being its Hermitian conjugate, 
$\gamma_{\mu} = (\gamma_0, \vec\gamma)$ are the Dirac matrices and $x\equiv (\vec{r},\,t)$ is the coordinate 4-vector ($\vec{r},\, t$ are the space- and time-coordinates). The photon field, or the photon wave function $A_{\mu}(x)$ looks like
\begin{eqnarray}
\label{2}
A^{(\vec e,\,\vec k)}_{\mu}(x) = \sqrt{\frac{2\pi}{\omega}}\,e^{(\lambda)}_{\mu}e^{i(\vec{k}\vec{r}-\omega t)}
 = \sqrt{\frac{2\pi}{\omega}}e^{-i\omega t}\,A^{(\vec e,\,\vec k)}_{\mu}(\vec r\,)
\, ,
\end{eqnarray}
where $e^{(\lambda)}_{\mu}$ is the photon polarization 4-vector, $k=(\vec{k},\omega)$ is the photon momentum 4-vector ($\vec{k}$ is the wave vector, $\omega=|\vec{k}|$ is the photon frequency).

The transition amplitude $U_{A'A}$ is defined as
\begin{eqnarray}
\label{3}
\langle A'|\hat{S}^{(1)}|A\rangle = -2\pi\, i\delta\left(\omega-E_A+E_{A'}\right)U_{A'A}^{(1)}\, .
\end{eqnarray}
Transition probability per time unit (transition rate) is defined via $U_{A'A}$ as \cite{LKD}
\begin{eqnarray}
\label{4}
W_{A'A}= 2\pi\left|U_{A'A}^{(1)}\right|^2\delta\left(\omega-E_A+E_{A'}\right)\, .
\end{eqnarray}

If the final state belongs to the continuous spectrum (as in our case due to the emitted photon) the differential transition probability should be introduced:
\begin{eqnarray}
\label{6}
dW_{A'A}(\vec{k},\vec{e})=2\pi\left|U_{A'A}^{(1)}\right|^2\delta\left(\omega-E_A+E_{A'}\right)
\frac{d\vec k}{(2\pi)^3}\, ,
\end{eqnarray}
where $d\vec k\equiv d^3k = \omega^2 d\vec{\nu}d\omega $, $d\vec{\nu}$ is the element of the solid angle in the momentum space. 
Integration in Eq. (\ref{6}) over $\omega$ gives the probability of the photon emission with polarization $\vec{e}$ in the direction $\vec{\nu}\equiv\vec{k}/\omega$ per time unit within solid angle $d\vec{\nu}$:
\begin{eqnarray}\label{7}
dW_{A'A}=\frac{e^2}{2\pi}\omega_{A'A}\left|\left(( \vec{e}^{\,*}\vec{\alpha})e^{-i\vec{k}\vec{r}}\right)_{A'A}\right|^2d\vec{\nu}\, ,
\end{eqnarray}
where $\omega_{A'A}=E_A-E_{A'}$. The total transition probability follows from Eq. (\ref{7}) after integration over angles and summation over the polarizations
\begin{eqnarray}
\label{8}
W_{A'A}=\frac{e^2}{2\pi}\omega_{A'A}\sum\limits_{\vec{e}}\int d\vec{\nu}\left|\left(( \vec{e}^{\,*}\vec{\alpha})e^{-i\vec{k}\vec{r}}\right)_{A'A}\right|^2
\end{eqnarray}

For the atomic electron the characteristic scales for $|\vec{r}|$ and $|\vec{k}|=\omega$ are: $|\vec{r}|\sim 1/m\alpha Z$, $\omega=E_{A'}-E_A\sim m(\alpha Z)^2$. Then in the nonrelativistic case, in particular for the hydrogen atom ($Z=1$), $\vec{k}\vec{r}\sim \alpha$ and the exponential function in the matrix element in Eq. (\ref{8}) can be replaced by 1. In the nonrelativistic limit the matrix element involving the Dirac matrices $\vec{\alpha}$ (electron velocity operator in the relativistic theory) can be substituted by the matrix element of the operator $\hat{\vec{p}}/m$; where $\hat{\vec{p}}$ is the electron momentum operator, with the Schr\"{o}dinger wave functions. Then Eq. (\ref{8}) takes the form
\begin{eqnarray}
\label{9}
W_{A'A}=\frac{e^2}{2\pi m^2}\omega_{A'A}\sum\limits_{\vec{e}}\int d\vec{\nu}\left|\left(\vec{e}\vec{p}\right)_{A'A}\right|^2\, ,
\end{eqnarray}
where the notation $(...)_{A'A}$ now implies evaluation of the matrix element with Schr\"{o}dinger 
wave functions. Performing summation over the polarization with the help of the standard formulas \cite{LKD} and integrating over $\vec{\nu}$ yields
\begin{eqnarray}
\label{10}
W_{A'A}^{\upsilon}=\frac{4}{3}\frac{e^2}{m^2}\omega_{A'A}\left|(\vec{p})_{A'A}\right|^2\, .
\end{eqnarray}
This is transition probability in the nonrelativistic limit in the "velocity" form. The "length" form $W^l_{A'A}$ involving the electric dipole moment operator $\vec{d}=e\vec{r}$ of the electron can be obtained from Eq. (\ref{10}) via the quantum mechanical relation
\begin{eqnarray}
\label{11}
\omega_{A'A}(\vec{r})_{A'A}=\frac{i}{m}(\vec{p})_{A'A}\, .
\end{eqnarray}
Then
\begin{eqnarray}
\label{12}
W_{A'A}^{l}=\frac{4}{3}\omega_{A'A}^3\left|(\vec{d})_{A'A}\right|^2\, .
\end{eqnarray}
Thus in the nonrelativistic limit only the electric dipole (E1) photon emission is allowed. Using the atomic characteristic scales for $|\vec{r}|$ and $\omega$, given above we easily obtain the order-of-magnitude estimates for the one-photon E1 transitions $W_{E1}^{(1\gamma)}$ mentioned in the previous Section. In particular for $2p-1s$ transition ($A=2p$, $A'=1s$) it follows from Eq. (\ref{12})
\begin{eqnarray}\label{13}
W_{E1}^{(1\gamma)}(2p-1s)=732.722 m\alpha(\alpha Z)^4 = 0.626\cdot 10^9\,\,\, s^{-1}
\end{eqnarray}

\section{QED derivation of the Lorentz profile.}

In this Section we give the QED description of the line profile. The basic ideas of the modern QED line profile theory were formulated by Low \cite{Low}.This method can be applied also for the description of the cascades and two-photon transitions. We employ the relativistic units ($\hslash=c=1$) throughout this section.

\subsection{Resonant scattering of the photon on the atomic electron.}

Consider first the elastic photon scattering process. The Fyenman graph corresponding to the process is presented in Fig. 3.

According to the standard correspondence rules for the bound-electron QED (see, for example, \cite{LKD}) the S-matrix element, corresponding to the graph Fig. 3, is
\begin{eqnarray}
\label{18}
\langle A'|\hat{S}^{(2)}|A\rangle  = e^2\int d^4x_1d^4x_2\left(\bar{\psi}_{A'}(x_1)\gamma_{\mu_1}A^*_{\mu_1}(x_1)S(x_1x_2)\gamma_{\mu_2}A^*_{\mu_2}(x_2)\psi_A(x_2)\right),
\end{eqnarray}
where $S(x_1x_2)$ is the Feynman propagator for the atomic electron. 
In the Furry picture the eigenmode decomposition for this propagator reads (e.g. \cite{Akhiezer})
\begin{eqnarray}
\label{19}
S(x_1x_2)=\frac{1}{2\pi i}\int\limits_{-\infty}^{\infty}d\omega_1e^{i\omega_1(t_1-t_2)}\sum\limits_n\frac{\psi_n(\vec{r}_1)\bar{\psi}_n(\vec{r}_2)}{E_n(1-i0)+\omega_1}\, ,
\end{eqnarray}
where the summation in Eq. (\ref{19}) extends over the entire Dirac spectrum of electron states $n$ in the field of the nucleus.

Inserting the expression for the propagator in the S-matrix element, integrating over the time variables and frequency variable $\omega_1$ and using the connection between the $S$-matrix and the amplitude $U_{A'A}$ Eq. (\ref{3}), we obtain an expression for the scattering amplitude
\begin{eqnarray}
\label{scat}
U_{sc}^{(2)}=\sum\limits_n\frac{\left(U^*_{\omega}\right)_{An}\left(U_{\omega'}\right)_{nA}}{E_n-E_A-\omega}
\end{eqnarray}
with condition $\omega=\omega'$ which implies the energy conservation. Here
$U_{\omega}\equiv e\gamma_{\mu}A_{\mu}(x)$ and $\omega$ denotes the frequency of the photon. In the resonance approximation the photon frequency $\omega$ is close to the difference of the two atomic levels: $\omega=\omega_{res}\approx E_{A'}-E_A$. Then we can retain only one term in the sum over $n$ in Eq. (\ref{scat}):
\begin{eqnarray}
\label{res1}
U_{sc}^{(2)res}=\frac{\left(U^*_{\omega}\right)_{AA'}\left(U_{\omega'}\right)_{A'A}}{E_{A'}-E_A-\omega}\, .
\end{eqnarray}

\subsection{Line profile for the emission process.}

It follows from Eq. (\ref{res1}) that in the resonance approximation the emission amplitude can be expressed like
\begin{eqnarray}
\label{lp1}
U_{em}=\frac{\left(U^*_{\omega}\right)_{AA'}}{E_{A'}-E_A-\omega}\, .
\end{eqnarray}
The absorbtion amplitude can be presented similarly. Expression (\ref{res1}) for the scattering amplitude is singular at the resonant frequency. To remove this singularity one has to consider the electron self-energy insertion in the electron propagator in Fig. 3. According to \cite{Low} this implies the arrival of the Lorentz line profile for the scattering process. The lowest-order electron self-energy insertion is shown in Fig. 4. Using the correspondence rules, we obtain an expression for the correction to the scattering amplitude
\begin{eqnarray}
\label{lp2}
U_{sc}^{(4)}=-\sum\limits_{n_1\, n_2}\frac{\left(U^*_{\omega}\right)_{An_1}\left[\hat{\Sigma}(E_A+\omega)\right]_{n_1\, n_2}\left(U_{\omega'}\right)_{n_2A}}{(E_{n_1}-E_A-\omega)(E_{n_2}-E_A-\omega)}\, ,
\end{eqnarray}
where $\hat{\Sigma}(E)$ is the electron self-energy operator for the bound electron \cite{LKD}.

In the resonance approximation $n_1=n_2=A'$ and the correction to the scattering amplitude is 
\begin{eqnarray}
\label{lp3}
U_{sc}^{(4)}=-U_{sc}^{(2)res}\frac{\left(\hat{\Sigma}(E_A+\omega)\right)_{A'\, A'}}{E_{A'}-E_A-\omega}\, .
\end{eqnarray}
Repeating the insertions in the resonance approximation we obtain the geometric progression. The summation of this progression yields 
\begin{eqnarray}
\label{lp4}
U_{sc}^{res}=\frac{\left(U^*_{\omega}\right)_{AA'}\left(U_{\omega'}\right)_{A'A}}{\tilde{E}_{A'}-E_A-\omega}\, ,
\end{eqnarray}
where $\tilde{E}_{A'}=E_{A'}+\left(\hat{\Sigma}(E_A+\omega)\right)_{A'\, A'}$. The emission amplitude in the resonance approximation is then presented by an expression
\begin{eqnarray}
\label{lp5}
U_{em}=\frac{\left(U^*_{\omega}\right)_{AA'}}{\tilde{E}_{A'}-E_A-\omega}\, .
\end{eqnarray}
Apart from the electron self-energy (SE) also the vacuum polarization (VP) insertion in the electron propagator in Fig. 3 should be considered to all orders in the resonance approximation. The lowest-order VP insertion is described in Fig. 5. The VP insertions lead to the following change of the nenergy denominator:
\begin{eqnarray}
\label{lp6}
\tilde{E}_{A'}=E_A+\left(\hat{\Sigma}(E_A+\omega)\right)_{A'\, A'}+\left(\hat{\Pi}\right)_{A'\, A'}\, ,
\end{eqnarray}
where $(\hat{\Pi})_{A'\, A'}$ is the vacumm polarization operator for the bound electron \cite{LKD}.

The real part of the matrix element $\left(\hat{\Sigma}(E_A+\omega)\right)_{A'\, A'}$ presents the lowest order contribution of the electron self-energy to the Lanb shift, the imaginary part of this matrix element defines the radiative width $\Gamma_{A'}$ of the level $A'$:
\begin{eqnarray}
\label{lp7}
\left(\hat{\Sigma}(E_A+\omega)\right)_{A'\, A'}=L_{A'}^{SE}-\frac{i}{2}\Gamma_{A'}\, .
\end{eqnarray}
The other lowest-order part of the Lamb shift is the vacuum polarization part
\begin{eqnarray}
\label{lp8}
(\hat{\Pi})_{A'\, A'}=L_{A'}^{VP}\, .
\end{eqnarray}
The vacuum polarization does not contribute to the width. Then 
\begin{eqnarray}
\label{lp9}
\tilde{E}_{A'}=E_{A'}+L_{A'}-\frac{i}{2}\Gamma_{A'}\, ,
\end{eqnarray}
where $L_{A'}=L_{A'}^{SE}+L_{A'}^{VP}$ and the emission amplitude looks like
\begin{eqnarray}
\label{lp10}
U_{em}=\frac{\left(U^*_{\omega}\right)_{AA'}}{E_{A'}+L_{A'}-E_A-\omega-\frac{i}{2}\Gamma_{A'}}\, .
\end{eqnarray}
The total transition probability $A'\rightarrow A$ is
\begin{eqnarray}
\label{lp11}
dW_{AA'}(\omega)=\frac{1}{2\pi}\sum\limits_{\vec{e}}\int d\vec{\nu}\left|U_{em}\right|^2\omega^2d\omega\, .
\end{eqnarray}
Insertion of the expression (\ref{lp10}) in Eq. (\ref{lp11}) in the resonance approximation yields
\begin{eqnarray}
\label{lp12}
dW_{AA'}(\omega)=\frac{1}{2\pi}\frac{\Gamma_{AA'}d\omega}{(E_{A'}+L_{A'}-E_A-\omega)^2+\frac{1}{4}\Gamma_{A'}^2}\, .
\end{eqnarray}
Here $\Gamma_{AA'}$ is the partial width of the level $A'$ connected with the transition $A'\rightarrow A$.
Equation (\ref{lp12}) defines the usual Lorentz profile for the emission spectral line. The resonance frequency in zero-order approximation is defined by $\omega=\omega_{res}\approx E_{A'}-E_A$.

\section{Two-photon decay rate for the $2s-1s$ transition.}

In this section we discuss the two-photon processes $2s\rightarrow 1s + 2\gamma(E1)$ which corresponds to the decay scheme Fig. 6. The two-photon transition probability $A\rightarrow A'+2\gamma$ corresponds to the second-order $S$-matrix element Eq. (\ref{18}) (see Fig. 7) with the replacement of the absorbed photon by another emitted one. Using again Eqs. (\ref{3}) and (\ref{4}) for the two-photon transition and integrating over time and frequency variables in Eq. (\ref{18}), we find for the sum of the contributions of the both Feynamn graphs (see Figs. 7)
\begin{eqnarray}
\label{20}
dW_{A'A}=2\pi\delta\left(E_A-E_{A'}-\omega-\omega'\right)\left|U_{A'A}^{(2)}\right|^2\frac{d\vec{k}}{(2\pi)^3}\frac{d\vec{k}'}{(2\pi)^3}\, ,
\end{eqnarray}
where 
\begin{eqnarray}
\label{21}
U_{A'A}^{(2)} = \frac{2\pi e^2}{\sqrt{\omega\omega'}}\left[\sum\limits_n\frac{\left(\vec{\alpha}\vec{A}^*_{\vec{e},\vec{k}}\right)_{A'n}\left(\vec{\alpha}\vec{A}^*_{ \vec{e}\,',\vec{k}'}\right)_{nA}}{E_n-E_A+\omega'}+\sum\limits_n\frac{\left(\vec{\alpha}\vec{A}^*_{ \vec{e}\,',\vec{k}'}\right)_{A'n}\left(\vec{\alpha}\vec{A}^*_{\vec{e},\vec{k}}\right)_{nA}}{E_n-E_A+\omega}\right]\, ,
\end{eqnarray}
$\vec{A}_{\vec{e}, \vec{k}} = \vec{e}\,e^{i\vec{k}\vec{r}}$ and $\vec{\alpha}$ are the Dirac matrices..

In what follows, we will be interested in the decay rate of the $ns$ levels 
($A\equiv ns$, $A'\equiv 1s$)
in hydrogen. In this section we focus on the case $n=2$, when the cascades are absent. The schematic picture of the decay $2s\rightarrow 1s+2\gamma(E1)$ is given in Fig. 6. In the nonrelativistic limit, after the integration over frequency $\omega'$, over photon directions 
$d\vec{\nu}$, $d\vec{\nu}'$ and summation over all  polarizations $\vec{e}$, $ \vec{e}\,'$, we obtain for the photon frequency distribution:  
\begin{eqnarray}
\label{23}
dW_{2s,1s}(\omega)=\frac{8\omega^3(\omega_0-\omega)^3}{27\pi}e^4\left|S_{1s,2s}(\omega)+S_{1s,2s}(\omega_0-\omega)\right|^2d\omega\, ,
\end{eqnarray}
\begin{eqnarray}
\label{24}
S_{1s,2s}(\omega)=\sum\limits_{n'p}\frac{\langle R_{1s}|r|R_{n'p}\rangle\langle R_{n'p}|r|R_{2s}\rangle}{E_{n'p}-E_{ns}+\omega}\, ,
\end{eqnarray}
\begin{eqnarray}
\label{25}
\langle R_{n'l'}|r|R_{nl}\rangle=\int\limits_{0}^{\infty}r^3R_{n'l'}(r)R_{nl}(r)dr\, ,
\end{eqnarray}
where $\omega_0=E_{2s}-E_{1s}$, $R_{nl}(r)$ are the radial part of the nonrelativistic hydrogen wave functions, and $E_{nl}$ are the hydrogen electron energies. Here we have used again the quantum-mechanical relation Eq. (\ref{11}); Eq. (\ref{23}) is written in the "length" form.

The decay rate for the two-photon transition can be obtained by integration of Eq. (\ref{23}) over the entire frequency interval 
\begin{eqnarray}
\label{26}
W_{2s,1s}=\frac{1}{2}\int\limits_0^{\omega_0}dW_{2s,1s}(\omega).
\end{eqnarray}
In case of $2s$ state the cascade transitions are absent, the frequency distribution Eq. (\ref{23}) is not singular and the integral Eq. (\ref{26}) is convergent. The result of the integration over frequency $\omega$ is well known and equal to $W_{2s,1s}=24.788m\alpha^2(\alpha Z)^6\, r.u. = 8.229 \, s^{-1}$ \cite{Klarsfeld}.

\section{Two-photon decay with cascades for the $3s-1s$ transition.}

In case of the cascade transitions ($n> 2$), some terms in Eq. (\ref{24}) become singular and the integral Eq. (\ref{26}) diverges. This divergency has a physical origin: an emitted photon meets the resonance. The corresponding scheme of the decay for $n=3$ is given in Fig. 8. So the divergency can be avoided only by introducing the width of this resonance.This situation was studied in \cite{LabShon} for the HCI. The same recipe can also be used in case of the hydrogen atom. Following the prescriptions given in \cite{LabShon} we separate out the resonant terms (corresponding to cascades) in the sum over the intermediate states Eq. (\ref{24}) and apply Low's procedure \cite{Low} for the regularization of the corresponding expressions in the vicinity of the resonance frequency values. Practically this 
leads to the apperance of the energy level widths in the energy denominators. Then the Lorentz profiles arise for the resonant terms in the expression for the probability. However, the Lorentz profile is valid only in the vicinity of the resonance and cannot be extended too far off 
from the resonance frequency value. As for any multichannel processes such a separation is an approximate procedure due to existence of the interference terms.

The integration over the entire frequency interval $[0,\omega_0]$ in Eq. (\ref{26}) should be split into several subintervals, e.g. 5 in case of the two-photon emission profile for the $3s$-level decay, see Fig. 9 \cite{LSP}. The first interval (I) extends from $\omega=0$ up to the lower boundary of the second interval (II). The latter one encloses the resonance frequency value $\omega_1=E_{3s}-E_{2p}$. Within the interval (II) the resonant term $n=2$ in Eq. (\ref{24}) should be subtracted from the sum over intermediate states and replaced by the term with modified energy denominator (see Section IV). This modified denominator is $E_{2p}-E_{3s}+\omega+\frac{i}{2}\Gamma$, where $\Gamma=\Gamma_{2p}+\Gamma_{3s}$. The third interval (III) extends from the upper boundary of interval II up to the lower boundary of the interval (IV), the latter one enclosing another resonance frequency value $\omega_2=E_{2p}-E_{1s}$. Within the interval (IV) again the resonant term $n=2$ in Eq. (\ref{24}) should be replaced by the term with modified denominator $E_{2p}-E_{1s}-\omega-\frac{i}{2}\Gamma_{2p}$. Finally, a fifth interval (V) ranges from the upper boundary of the interval (IV) up to the maximum frequency value $\omega_0$. Note, that the frequency distribution $dW_{3s,1s}(\omega)$ is symmetric with respect to $\omega=\omega_0/2$ with a 1\% accuracy (the asymmetry is due to the difference between $\Gamma=\Gamma_{2p}+\Gamma_{2s}$ and $\Gamma_{2p}$, respectively).

Inserting Eq. (\ref{23}) into Eq. (\ref{26}) and retaining only the resonant term within the second and fourth frequency intervals, will yield the cascade contribution to the total two-photon decay rate of the $3s$-level. Taking the ratio to the total width of the $3s$-level $\Gamma_{3s}$ we will obtain the absolute probability or branching ratio $W^{(cascade)}_{3s;1s}/\Gamma_{3s}\equiv b^{(cascade)}_{3s-2p-1s}$ for the cascade transition. The contributions to $b^{(cascade)}_{3s-2p-1s}$ from the intervals (I), (III), (V) are assumed to be zero. The cascade contribution of the $3s$-level results (in the "length" form)
\begin{eqnarray}
\label{68}
W^{({\rm cascade}\, 1\gamma)}_{3s;1s}=\frac{4}{27\pi}\int\limits_{({\bf  II})}\omega^3(\omega_0-\omega)^3\left|\frac{\langle R_{3s}(r)|r|R_{2p}(r)\rangle\langle R_{2p}(r')|r'|R_{1s}(r')\rangle}{E_{2p}-E_{3s}+\omega-\frac{i}{2}\Gamma}\right|^2d\omega + 
\\
\nonumber
+\frac{4}{27\pi}\int\limits_{({\bf  IV})}\omega^3(\omega_0-\omega)^3\left|\frac{\langle R_{3s}(r)|r|R_{2p}(r)\rangle\langle R_{2p}(r')|r'|R_{1s}(r')\rangle}{E_{2p}-E_{1s}-\omega-\frac{i}{2}\Gamma_{2p}}\right|^2d\omega .
\end{eqnarray}

According to the discussion in Section V the "pure" two-photon decay probabilities within each interval, defined in Section V, look like
\begin{eqnarray}
\label{69}
dW_{3s;1s}^{(\rm{pure} 2\gamma)}&=&\frac{4}{27\pi}\omega^3(\omega_0-\omega)^3\left|S_{1s;3s}^{(2p)}(\omega)+S_{1s;3s}(\omega_0-\omega)\right|^2d\omega, \,\, \omega\in {\bf II}
\end{eqnarray}
\begin{eqnarray}
\label{70}
dW_{3s;1s}^{(\rm{pure} 2\gamma)}&=&\frac{4}{27\pi}\omega^3(\omega_0-\omega)^3\left|S_{1s;3s}(\omega)+S_{1s;3s}^{(2p)}(\omega_0-\omega)\right|^2d\omega, \,\, \omega\in {\bf IV}
\end{eqnarray}
\begin{eqnarray}
\label{71}
dW^{(\rm{pure} 2\gamma)}_{3s;1s} &=& \frac{4}{27\pi}\omega^3(\omega_0-\omega)^3\left|S_{1s;3s}(\omega)+S_{1s;3s}(\omega_0-\omega)\right|^2d\omega, \,\, \omega \in {\bf I, III, V} \, .
\end{eqnarray}
Here $S_{1s;3s}^{(2p)}(\omega)$ is the expression (\ref{24}) with the $n=2$ term 
being excluded.

Unlike cascade, all the intervals contribute to the "pure" two-photon transition. The branching ratio for this transition $3s\rightarrow 2\gamma +1s$ appears to be
\begin{eqnarray}
\label{72}
b^{(\rm{pure} 2\gamma)}_{3s-1s} = \frac{1}{2}\frac{1}{\Gamma_{3s}}\int\limits_0^{\omega_0}dW^{(pure 2\gamma)}_{3s;1s}(\omega)\, .
\end{eqnarray}
It remains to introduce the interference contribution. This contribution comes only from the intervals II and IV. The corresponding frequency distribution functions are given by 
\begin{eqnarray}
\label{73}
dW^{(\rm{inter})}_{3s;1s}=\frac{4\omega^3(\omega_0-\omega)^3}{27\pi}Re\left[\frac{\langle R_{3s}(r)|r|R_{2p}(2r)\rangle\langle R_{2p}(r')|r'|R_{1s}(r')\rangle}{E_{2p}-E_{3s}+\omega-\frac{i}{2}\Gamma_{2p}}\right]\left[S_{1s;3s}^{(2p)}(\omega)+S_{1s;3s}(\omega_0-\omega)\right]d\omega\,,\,\, \omega\in\, {\bf II}
\end{eqnarray}
\begin{eqnarray}
\label{74}
dW^{(\rm{inter})}_{3s;1s}=\frac{4\omega^3(\omega_0-\omega)^3}{27\pi}Re\left[\frac{\langle R_{3s}(r)|r|R_{2p}(2r)\rangle\langle R_{2p}(r')|r'|R_{1s}(r')\rangle}{E_{2p}-E_{1s}-\omega-\frac{i}{2}\Gamma_{2p}}\right]\left[S_{1s;3s}(\omega)+S_{1s;3s}^{(2p)}(\omega_0-\omega)\right]d\omega\,,\,\, \omega\in\, {\bf IV} 
\end{eqnarray}
and branching ratio results as 
\begin{eqnarray}
\label{75}
b^{(\rm{inter})}_{3s;1s} = \frac{1}{2\Gamma_{3s}}\int\limits_{({\bf II})}dW^{(\rm{inter})1}_{3s;1s}+\frac{1}{2\Gamma_{3s}}\int\limits_{({\bf IV})}dW^{(\rm{inter})2}_{3s;1s}.
\end{eqnarray}

The results of our calculations are presented in Table 1. 
It is convenient to define the size $\Delta \omega$ of the second interval as 
multiples $l$ of the widths $\Gamma$, i.e.   
$\Delta \omega = 2l\Gamma$ and for the fourth interval 
as $\Delta \omega = 2l\Gamma_{2p}$, respectively. In Table 1 numbers are given for different values of $l$ ranging from 
$l\simeq 10^5$ up to $l\simeq 10^7$. The upper boundary of interval {\bf II} equals  $\omega_1+l\Gamma=\frac{5}{72}+l\Gamma$ (in a.u.), while the lower boundary of interval 
{\bf IV} equals $\omega_2-l\Gamma_{2p}=\frac{3}{8}-l\Gamma_{2p}$. 
The different lines of the Table 1 present branching ratios and transition rates 
of the "pure" two-photon and "interference" channels, respectively. For the more detailed analysis the contributions of the "pure" two-photon transition rate for the each frequency interval are also compiled. The branching ratio and the transition rate for the cascade contribution can be obtained from the relation  $b^{(\rm{cascade})}_{3s-2p-1s} + b^{(\rm{pure}2\gamma)}_{2s;1s} + b^{(\rm{inter})}_{3s;1s}=1$. This relation is sutisfied with high accuracy since the only decay channel neglected is the very weak direct 1-photon $M1$ transition $3s\rightarrow 1s+\gamma$. From the Table 1 we can draw the same conlusions: as in the case of the HCI \cite{LabShon}: the "pure" two-photon and cascade contributions to the total decay rate appear to be inseparable. Changing the interval size $\Delta\omega$, 
we obtain quite different values for $dW^{(\rm{pure} 2\gamma)}_{3s;1s}$ ranging
from $202.16\, s^{-1}$ (for $l = 10^4$) up to $7.9385\, s^{-1}$ 
(for $l = 1.00256\cdot 10^7$).

Moreover, in our calculations - depending on the size of the interval - the interference contribution also can become quite large, comparable in magnitude with the "pure" two-photon
contribution. Thus, we demonstrated that even the 
order of magnitude of the "pure" two-photon decay rate for the $3s$-state in hydrogen can not be predicted reliably.

Earlier the result $8.2196$ $s^{-1}$ for the "pure" two-photon decay of the  $3s$-level was reported in \cite{cea86} and confirmed in \cite{fsm88}. However, as it was pointed out in \cite{J.Chluba} in both papers \cite{cea86}, \cite{fsm88} the summation over the intermediate states was not performed properly. The "nonresonant" contribution $10.556$ $s^{-1}$ deduced in \cite{J.Chluba}, which plays the role of the "pure" two-photon decay rate is well within the range of our values given Table 1. However, the result $2.08$ $s^{-1}$ obtained for the "pure" two-photon decay rate in \cite{jas08} is in strong contradiction with the present analysis (see the discussion in \cite{LSP}).

Very recently, a paper \cite{Amaro} did arrive where both the standard QED
approach, based on the line profile theory (\cite{Drake}-\cite{LabShon}) and the
"alternative" approach based on the two-loop Lamb shift theory (\cite{Jent1}-\cite{Jent3})
were applied to the calculation of the two-photon transition in hydrogen.
A reasonable agreement between the two methods was found. However, from
the derivations in \cite{LSP} it follows that the employment of the
Lamb shift imaginary part gives exactly the same results as the LPA
QED approach.  
\begin{table}[h]
\caption{
Branching ratios and transition rates (in $s^{-1}$) for the different decay channels for the decay probability of the $3s$ level with different frequency interval size ($l$).}
\small{
\begin{tabular}{| l | c | c | c | c | c | c | c | c |}
\hline \hline
$l $& $10^4$ & $10^{5}$&$2.5\cdot 10^5$& $5\cdot 10^5$ & $10^6$ & $1.5\cdot 10^6 $ &$4.53\cdot 10^6$& $1.00256\cdot 10^7$\\ \hline
$b^{(\rm{pure} 2\gamma)}$& $3.2003\cdot 10^{-5}$ & $3.5091\cdot 10^{-6}$&$1.6270\cdot 10^{-6}$&$1.0239\cdot 10^{-6}$ & $7.6765\cdot 10^{-7}$ & $ 7.2201\cdot 10^{-7}$ & $9.1487\cdot 10^{-6}$ &$1.2567\cdot 10^{-6}$\\ \hline
$W^{(\rm{pure} 2\gamma)}_{I}$ & $ 53.054$ & $ 7.0547$ &$ 3.5743$& $ 2.1898$ & $ 1.27737$ &$0.85130 $ & $2.4979\cdot 10^{-6} $& $0$\\ \hline
$W^{(\rm{pure} 2\gamma)}_{II}$ & $ 0.006247$ & $ 0.06247$ &$ 0.15614$& $ 0.31201$ & $ 0.62183$ &$0.92718 $ & $ 2.4666$& $ 3.9810$\\ \hline
$W^{(\rm{pure} 2\gamma)}_{III}$ & $ 95.536$ & $ 7.8778$ &$ 2.7928$& $ 1.4517$ & $ 1.0457$ &$1.0031 $ & $ 0.86005$& $0$\\ \hline
$W^{(\rm{pure} 2\gamma)}_{IV}$ & $ 0.006185$ & $ 0.061847$ &$0.15458$& $ 0.30890$ & $0.61569$ &$0.91813 $ & $ 2.4523$& $ 3.9575$\\ \hline
$W^{(\rm{pure} 2\gamma)}_{V}$ & $ 53.561$ & $ 7.1101$ &$ 3.5999$& $ 2.2056$ & $ 1.2886$ &$0.861254 $ & $ 3.1665\cdot 10^{-4}$& $0$\\ \hline
$W^{(\rm{pure} 2\gamma)} $ & $202.16$ & $22.167$ &$10.278$& $6.4680$ & $4.8492$ &$4.5609$ & $5.7792$& $7.9385$\\ \hline 
$b^{(\rm{inter})} $ & $-1.4342\cdot 10^{-9}$ & $-1.4343\cdot 10^{-8}$& $-3.5852\cdot 10^{-8}$& $-7.1665\cdot 10^{-8}$ & $-1.4302\cdot10^{-7}$ &$ -2.1376\cdot 10^{-7}$ & $-6.0829\cdot 10^{-7}$ & $-1.0459\cdot 10^{-6}$\\ \hline
$W^{(\rm{inter})} $ & $-0.0090599$ & $-0.090602$& $-0.22647$& $-0.45270$ & $-0.90346$ &$-1.3503$ & $-3.8426$ &$-6.6067$\\
\hline \hline
\end{tabular}
}
\end{table}

In the end of this Section we will explain why the contributions of the "pure" two-photon transition rates in Table 1 are of the same order as the interference terms. As it was mentioned earlier the contribution of the cascade is of the order $m\alpha (\alpha Z)^4$ in r.u. However, this is the result of the integration over the frequency interval of the order $\Gamma^{(1\gamma)}\sim m\alpha (\alpha Z)^4$. Then the order of the magnitude of the "amplitude" in the integrand is 1 (and dimensionless). The order of the magnitude of the "pure" two-photon contribution is $m\alpha^2(\alpha Z)^6$ r.u. This result also incorporates the integral over frequency interval of the order $\omega_0=m(\alpha Z)^2$ in r.u. (see Eq. (34)). Then the order of the magnitude of the corresponding "amplitude" in the integrand is $[m\alpha^2(\alpha Z)^6/m(\alpha Z)^2]^{1/2}=\alpha (\alpha Z)^2$. The latter value is again dimensionless. Multiplying the cascade and "pure" two-photon "amplitudes" in the integrand we will have the dimensionless integrand of the order $1\cdot \alpha (\alpha Z)^2$. Finally, integrating the product over the frequency interval $\Gamma\sim m\alpha(\alpha Z)^4$, where the interference terms are nonzero, we obtain the contribution of the order $m\alpha^2(\alpha Z)^4$ r.u. This is parametrically the same as the order of magnitude for the "pure" two-photon contribution.

\section{"Two-photon approximation" for the tree-photon 3p-1s transition in hydrogen.}

The $3p-1s$ decay can occur either as one-photon or as three-photon process. These channels do not interfere due to the different number of photons in the final state. The one-photon decay $3p\rightarrow 1s+\gamma(E1)$ corresponds to the decay scheme Fig. 1, where the initial state $2p$ should be replaced by $3p$. The value $W^{(1\gamma)}_{E1}(3p-1s)$ can be obtained from the formula (11) when inserting there $A\equiv 3p$, $A'\equiv 1s$. The result is
\begin{eqnarray}
W_{E1}^{(1\gamma)}(3p-1s)= 195.613 m\alpha(\alpha Z)^4=1.67342\cdot 10^8 \, s^{-1}\, .
\end{eqnarray}

The schematic picture for the process $A\rightarrow A'+3\gamma(E1)$ is given in Fig. 10 and the corresponding Feynman graph is depicted in Fig. 11. The three-photon emission probability was evaluated in our work \cite{SLPS} for the $2p-1s$ three-photon transition in hydrogen. The parametric estimate can be easily found and is equal to $m\alpha^3(\alpha Z)^8$ in relativistic units. The value of the probability for this transition is $0.4946 m\alpha^3(\alpha Z)^8$ r.u.

The $2p\rightarrow 1s+3\gamma(E1)$ transition is "pure" 3-photon transition (no cascade are possible in this case)and, according to our general scheme (see Section II) should be neglected. For $3p\rightarrow 1s+3\gamma(E1)$ transition the two cascades should be taken into account: $3p\rightarrow 2s+\gamma(E1)\rightarrow 1s+3\gamma(E1)$ and $3p\rightarrow 2p+2\gamma(E1)\rightarrow 1s+3\gamma(E1)$. The contribution of these two cascades will be studied in this Section. It will be shown that the decay rates of these two cascade channels will be comparable with the "pure" two-photon contribution of the $3s\rightarrow 1s+2\gamma(E1)$ channel. Unlike the $3s\rightarrow 1s+2\gamma(E1)$ case, where the contribution of the "pure" two-photon decay is nonseparable from the cascade contribution, for $3p\rightarrow 1s+3\gamma(E1)$ decay only the cascade contribution should be taken into account. The reason is that the "pure" 3-photon contribution is beyond the accuracy of the "two-photon" approximation, adopted in this paper.

The $S$-matrix element for the 3-photon decay process $A\rightarrow A'+3\gamma$ is:
\begin{eqnarray}
\label{1.three}
S_{A'\,A}^{(3)}=(-i)^3\int d^4x_1d^4x_2d^4x_3 \bar{\psi}_{A'}(x_1)\left(\gamma_{\mu_1}A^{*\omega''}_{\mu_1}(x_1)\right)S(x_1x_2)\left(\gamma_{\mu_2}A^{*\omega'}_{\mu_2}(x_2)\right)S(x_2x_3)\left(\gamma_{\mu_3}A^{*\omega}_{\mu_3}(x_3)\right)\psi_A(x_3)\, ,
\end{eqnarray}
where the photon field $A^{*\omega}_{\mu}(x)$ is described by Eq. (\ref{2}) and $\omega$, $\omega'$ $\omega''$ denote the frequencies of the photons. The electron propagator $S(x_1x_2)$ and the electron wave functions $\bar{\psi}_{A'}(x_1)$, $\psi_A(x_3)$ are defined like in Section III:
\begin{eqnarray}
\label{2.three}
\bar{\psi}_{A'}(x_1)=\bar{\psi}_{A'}(\vec{r}_1)e^{iE_{A'}t_1};\qquad \psi_A(x_3)=\psi_A(\vec{r}_3)e^{-iE_At_3}.
\end{eqnarray}
Using Eqs (\ref{2}), (\ref{19}) and (\ref{2.three}) we can perform the time integration over time variables in Eq. (\ref{2.three})
\begin{eqnarray}
\label{3.three}
\int dt_1e^{i(E_{A'}+\omega''+\omega_1)t_1}=2\pi\delta\left(E_{A'}+\omega''+\omega_1\right),
\nonumber
\\
\int dt_2e^{i(\omega'-\omega_1+\omega_2)t_2}=2\pi\delta\left(\omega'-\omega_1+\omega_2\right),
\\
\nonumber
\int dt_3e^{i(\omega-\omega_2-E_{A})t_3}=2\pi\delta\left(\omega-\omega_2-E_{A}\right).
\end{eqnarray}
Then the frequency variables in the two energy denominators are $\omega_1=-E_{A'}-\omega''$, $\omega_2=\omega_1-\omega'=-E_{A}-\omega''-\omega'$. From these two equations follows $\omega+\omega'+\omega''=E_{A}-E_{A'}$, what is the energy conservation law for this process. 

Then after the integration over $\omega_1$ and $\omega_2$ the $S$-matrix element can be written in the form:
\begin{eqnarray}
\label{4.three}
\langle A'|S^{(3)}|A\rangle = (-1)^3e^3\int d^3r_1d^3r_2d^3r_3\bar{\psi}_{A'}(\vec{r}_1)(\vec{e''}\vec{\alpha}_1)\sqrt{\frac{2\pi}{\omega''}}e^{-i(\vec{k''}\vec{r}_1)}\sum\limits_{n_1}\frac{\psi_{n_1}(\vec{r}_1)\bar{\psi}_{n_1}(r_2)}{E_{n_1}(1-i0)-E_{A'}-\omega''}\times
\\
\nonumber
(\vec{e'}\vec{\alpha}_2)\sqrt{\frac{2\pi}{\omega'}}e^{-i(\vec{k'}\vec{r}_2)}\sum\limits_{n_2}\frac{\psi_{n_2}(\vec{r}_2)\bar{\psi}_{n_2}(r_3)}{E_{n_2}(1-i0)-E_{A'}-\omega''-\omega'} (\vec{e}\vec{\alpha}_3)\sqrt{\frac{2\pi}{\omega}}e^{-i(\vec{k}\vec{r}_3)}\psi_A(\vec{r}_3)\delta(E_{A'}-E_A+\omega+\omega'+\omega'').
\end{eqnarray}
Here $\vec{\alpha}_i$ ($i=1,2,3$) is the Dirac matrix, $\vec{k}$ is the wave vector of the corresponded photon ($|\vec{k}|=\omega$) and $\vec{e}$ is the polarization vector of the emitted photon.

The transition probability per time unit is defined via Eqs (\ref{3}), (\ref{4}) by
\begin{eqnarray}
\label{5.three}
W^{(3\gamma)}_{A'A}=2\pi\left|U^{(3)}_{A'A}\right|^2\delta(E_{A'}-E_A+\omega+\omega'+\omega'').
\end{eqnarray}

The differential transition probability is introduced like
\begin{eqnarray}
\label{6.three}
dW^{(3\gamma)}_{A'A}(\vec{k},\vec{e}; \vec{k'},\vec{e'}; \vec{k''},\vec{e''})2\pi\left|U^{(3)}_{A'A}\right|^2\delta(E_{A'}-E_A+\omega+\omega'+\omega'')\frac{d^3k}{(2\pi)^3}\frac{d^3k'}{(2\pi)^3}\frac{d^3k''}{(2\pi)^3}\, ,
\end{eqnarray}
where $d^3k\equiv \omega^2 d\omega d\vec{\nu}$ and $d\vec{\nu}$ is the element of the corresponding solid angle in the momentum space.

The total transition probability follows from Eq. (\ref{6.three}) after integration over angles and summation over polarizatoins
\begin{eqnarray}
\label{7.three}
W^{(3\gamma)}_{A'A}=\frac{e^6}{3! (2\pi)^5}\sum\limits_{\vec{e},\vec{e'},\vec{e''}}\int d\vec{\nu}\int d\vec{\nu'}\int d\vec{\nu''}\int \omega d\omega\int \omega'd\omega'\int \omega''d\omega''\times
\nonumber
\\
\left|\sum\limits_{n_1,n_2}\frac{\langle A'|(\vec{e''}\vec{\alpha})e^{-i(\vec{k''}\vec{r}_1)}|n_1\rangle \langle n_1|(\vec{e'}\vec{\alpha})e^{-i(\vec{k'}\vec{r}_2)}|n_2\rangle \langle n_2|(\vec{e}\vec{\alpha})e^{-i(\vec{k}\vec{r}_3)}|A \rangle}{(E_{n_1}-E_{A'}-\omega'')(E_{n_2}-E_{A'}-\omega'-\omega'')}+...\right|^2.
\end{eqnarray}
where $\vec{k}$, $\vec{e}$; $\vec{k'}$, $\vec{e'}$; $\vec{k''}$, $\vec{e''}$ are the wave vectors and the polarization vectors for the three emitted photons. In Eq. (\ref{7.three}) the permutation symmetry of the emitted photons is taken into account.

Consider now the $3p\rightarrow 3\gamma(E1)+1s$ transition. In this case the initial state $A=3p$ and final state is $A'=1s$. Due to the conservation law we can write $\omega'+\omega''=E_{3p}-E_{1s}-\omega$ and, therefore, the transition probability (in dipole approximation) is
\begin{eqnarray}
\label{8.three}
W_{1s,3p}^{(3\gamma)}=\frac{e^6}{3!(2\pi)^5}\sum\limits_{\vec{e},\vec{e'},\vec{e''}}\int d\vec{\nu}\int d\vec{\nu'}\int d\vec{\nu''}\int \omega d\omega\int \omega''d\omega'' (\omega_0-\omega-\omega'')\times 
\nonumber
\\
\left|\sum\limits_{n_1,n_2}\frac{\langle 1s|(\vec{e''}\vec{p})|n_1\rangle \langle n_1|(\vec{e'}\vec{p})|n_2\rangle \langle n_2|(\vec{e}\vec{p})|3p \rangle}{(E_{n_1}-E_{A'}-\omega'')(E_{n_2}-E_{A'}+\omega)}+...\right|^2\, ,
\end{eqnarray}
where $\omega_0 \equiv E_{3p}-E_{1s}$ and $\vec{p}$ is the electron momentum operator.

This three-photon emission probability contains resonant transition, when $E_{n_2}\equiv E_{2s}$. The value $E_{n_2} = E_{2s}$ corresponds to the pole in the integral over frequencies in Eq. (\ref{8.three}) $\omega=E_{3p}-E_{2s}$. Then $\omega_0-\omega-\omega''=E_{3p}-E_{1s}-E_{3p}+E_{2s}-\omega''=E_{2s}-E_{1s}-\omega''$ at the pole value. The integral over $\omega'$ becomes divergent. It can be regularized in the resonance region according to Section IV. After the regularization procedure we can integrate Eq. (\ref{8.three}) over frequency $\omega$:
\begin{eqnarray}
\label{9.three}
\sum\limits_{\vec{e}}\int d\vec{\nu}\int\omega d\omega\frac{\left|\langle 2s|(\vec{e}\vec{p})|3p\rangle\right|^2}{(E_{2s}-E_{3p}+\omega)^2+\frac{1}{4}\Gamma_{3p}^2}=\frac{\Gamma^{(1\gamma)}_{3p-2s}}{\Gamma_{3p}}2\pi^2.
\end{eqnarray}
The $\Gamma^{(1\gamma)}_{3p-2s}$ is the partial width corresponded to the $3p\rightarrow 1\gamma+2s$ transition and $\Gamma_{3p}$ is the total width of the 3p level which is the sum of the all partial widths for the transitions from 3p state to the ground state.

The remaining factor in Eq. (\ref{8.three}) represents the two-photon $2s-1s$ transition probability. Then the expression for the resonant contribution after collecting all the terms in the right-hand side of Eq. (\ref{8.three}) reads
\begin{eqnarray}
\label{10.three}
W^{(3\gamma)res1}_{3p-1s}=\frac{3}{4} W_{2s-1s}^{(2\gamma)}\frac{\Gamma^{(1\gamma)}_{3p-2s}}{\Gamma_{3p-1s}}.
\end{eqnarray}

In the sum over all intermediate states ($n_1$) in Eq. (\ref{8.three}) exists also another resonant term: $E_{n_1}\equiv E_{2p}$, when $\omega''=E_{2p}-E_{1s}$. In this case $\omega_0-\omega-\omega''=E_{3p}-E_{1s}-E_{2p}+E_{1s}-\omega=E_{3p}-E_{2p}-\omega$. Regularization in the region $\omega''=E_{3p}-E_{2p}$ leads to the final expression
%\begin{eqnarray}
%\label{11.three}
%W^{(3\gamma)}_{3p-1s}=\frac{e^6}{(2\pi)^5}\sum\limits_{\vec{e},\vec{e'},\vec{e''}}\int d\vec{\nu}\int d\vec{\nu'}\int d\vec{\nu''}\int \omega d\omega\int \omega''d\omega'' (\omega_0-\omega-\omega'')\times
%\\
%\frac{\left|\langle 1s|(\vec{e''}\vec{p}_1)| 2p\rangle\right|^2}{(E_{2p}-E_{1s}-\omega'')^2+\frac{1}{4}\Gamma_{2p}^2}\left|\sum\limits_{n_2}\frac{\langle 2p|(\vec{e'}\vec{p}_2)|n_2 \rangle\langle n_2|(\vec{e}\vec{p}_3)|\rangle}{E_{n_2}-E_{3p}+\omega}+...\right|^2.
%\end{eqnarray}
%The integral
%\begin{eqnarray}
%\label{12.three}
%\frac{1}{2\pi}\sum\limits_{\vec{e}}\int d\vec{\nu}\int \omega''d\omega''\frac{\left|\langle 1s|(\vec{e''}\vec{p}_1)| 2p\rangle \right|^2}{(E_{2p}-E_{1s}-\omega'')^2+\frac{1}{4}\Gamma^2_{2p}}=\int\limits_0^{E_{2p}-E_{1s}} \frac{\Gamma^{(1\gamma)}_{2p}d\omega''}{(E_{2p}-E_{1s}-\omega'')^2+\frac{1}{4}\Gamma^2_{2p}}=\pi
%\end{eqnarray}
%and, finally,
\begin{eqnarray}
\label{13.three}
W_{3p-1s}^{(3\gamma)res2}=\frac{3}{4} W_{3p-2p}^{(2\gamma)}.
\end{eqnarray}

Thus in the "two-photon" approximation the total probability of the $3p$ level decay can be expressed as
\begin{eqnarray}
\label{14.three}
W_{3p-1s}^{\rm{total}}=W_{3p-1s}^{(1\gamma)}+\frac{3}{4}W_{3p-2p}^{(2\gamma)}+\frac{3}{4}\frac{W_{3p-2s}^{(1\gamma)}}{\Gamma_{3p}}W_{2s-1s}^{(2\gamma)}
\end{eqnarray}

The first term in Eq. (\ref{14.three}) is the ordinary one-photon width. The two other terms represnt "cascade two-photon" contributions which are of the same order of magnitude as the "pure" two-photon contribution to the $3s$ level width $\Gamma_{3s}$. However, unlike the case of $3s$ level, where the "pure" two-photon contribution cannot be distinctly separated out \cite{LSP}, the "cascade two-photon" contribution to $\Gamma_{3p}$ is given explicitly by the two last terms in the right-hand side of Eq. (\ref{14.three}). The corresponding numerical values for hydrogen atom are
\begin{eqnarray}
\frac{3}{4}W_{3p-2p}^{(2\gamma)}=0.034005675\, s^{-1}
\nonumber
\\
\frac{3}{4}\frac{W_{3p-2s}^{(1\gamma)}}{\Gamma_{3p}}W_{2s-1s}^{(2\gamma)}=0.730334\, s^{-1}
%\\
%\nonumber
%\Gamma_{3p}=W_{3p-2s}^{(1\gamma)}+W_{3p-1s}^{(1\gamma)}+W_{3p-2p}^{(2\gamma)}=1.89803\cdot 10^8\, s^{-1}
\end{eqnarray}
which is not negligible in comparison with the order of magnitude of the two-photon contribution $\sim 10\, s^{-1}$ to the decay rate of $3s$ level. Evaluation of $W_{3p-2p}^{(2\gamma)}$ in Eq. (\ref{14.three}) is performed in the nonrelativistic limit, when the sublevels $2p_{1/2}$ and $2p_{3/2}$ are degenerate. The calculation of the transition rates between separate fine-structure components of $3p$, $2p$ levels is given in the Appendix A. The interference term between the second and the third terms in the right-hand side of Eq. (\ref{14.three}) are absent as well as the interference with the one-photon transition (the first term in the right-hand side of Eq. (\ref{14.three})). The reason is that the second and third terms are nonzero close to the two different frequency values $\omega$ in the integral in Eq. (51). Therefore the product of the two corresponding amplitudes in the integrand in Eq. (51) is always small.

\section{"Two-photon approximation" for the four-photon 4s-1s transition in hydrogen.}

Consider now the transition $4s\rightarrow 1s$. Apart from the ordinary two-photon processes $4s\rightarrow 3p + \gamma \rightarrow 1s+2\gamma$ and $4s\rightarrow 2p +\gamma\rightarrow 1s+2\gamma$ there are more complicated 4-photon processes with two-photon links. The schematic picture for the two-photon decays of $4s$ state is given in Fig. 12 and the 4-photon decay picture is presented in Fig. 13. In general, the S-matrix element for the 4-photon decay can be described by
\begin{eqnarray}
\label{1.four}
S_{4s, 1s}^{(4)}=(-i)^4\int d^4x_1d^4x_2d^4x_3 d^4x_4 \bar{\psi}_{A'}(x_1)\left(\gamma_{\mu_1}A^{*\omega'''}_{\mu_1}(x_1)\right)S(x_1x_2)\left(\gamma_{\mu_2}A^{*\omega''}_{\mu_2}(x_2)\right)\times
\\
\nonumber
S(x_2x_3)\left(\gamma_{\mu_3}A^{*\omega'}_{\mu_3}(x_3)\right)S(x_3x_4)\left(\gamma_{\mu_4}A^{*\omega}_{\mu_4}(x_4)\right)\psi_A(x_4)\, ,
\end{eqnarray}
where all the notations are the same as in Eq. (44). The electron wave functions $\bar{\psi}_{A'}(x_1)$, $\psi_A(x_4)$ are the same as in Eq. (45), with $A'=4s$, $A=1s$.
Perfoming time integration and integration over frequencies $\omega_1$, $\omega_2$, $\omega_3$, we recieve
\begin{eqnarray}
\label{3.four}
\langle A'|S^{(4)}|A\rangle = (-1)^4e^4\int d^3r_1d^3r_2d^3r_3d^3r_4\bar{\psi}_{A'}(\vec{r}_1)(\vec{e'''}\vec{\alpha}_1)\sqrt{\frac{2\pi}{\omega'''}}e^{-i(\vec{k'''}\vec{r}_1)}\sum\limits_{n_1}\frac{\psi_{n_1}(\vec{r}_1)\bar{\psi}_{n_1}(r_2)}{E_{n_1}(1-i0)-E_{A'}-\omega'''}\times
\nonumber
\\
(\vec{e''}\vec{\alpha}_2)\sqrt{\frac{2\pi}{\omega''}}e^{-i(\vec{k''}\vec{r}_2)}\sum\limits_{n_2}\frac{\psi_{n_2}(\vec{r}_2)\bar{\psi}_{n_2}(r_3)}{E_{n_2}(1-i0)-E_{A'}-\omega'''-\omega''} (\vec{e'}\vec{\alpha}_3)\sqrt{\frac{2\pi}{\omega'}}e^{-i(\vec{k'}\vec{r}_3)}\sum\limits_{n_3}\frac{\psi_{n_3}(\vec{r}_3)\bar{\psi}_{n_3}(r_4)}{E_{n_3}(1-i0)-E_{A'}-\omega''-\omega'}\times
\\
\nonumber
 (\vec{e}\vec{\alpha}_4)\sqrt{\frac{2\pi}{\omega}}e^{-i(\vec{k}\vec{r}_4)}\psi_A(\vec{r}_4)\delta(E_{A'}-E_A+\omega+\omega'+\omega''+\omega''').
\end{eqnarray}
Here $\vec{k}$, $\vec{e}$; $\vec{k'}$, $\vec{e'}$; $\vec{k''}$, $\vec{e''}$; $\vec{k'''}$, $\vec{e'''}$ are the wave vectors and polarization vectors for the four emitted photons. Using Eq. (\ref{3}) we can write the transition probability like
\begin{eqnarray}
\label{4.four}
W_{AA'}^{(4\gamma)}=\frac{e^8}{4!(2\pi)^7}\sum\limits_{\vec{e}\vec{e'}}\sum\limits_{\vec{e''}\vec{e'''}}\int d\vec{\nu}\int d\vec{\nu'}\int d\vec{\nu''}\int d\vec{\nu'''}\int\omega d\omega\int\omega' d\omega'\int\omega'' d\omega''\int\omega''' d\omega'''\left|U^{(4)}_{A'A}\right|^2\,.
\end{eqnarray}

Now we will demonstrate how the two-photon emission is included in this four-photon process. First, we consider the $4s\rightarrow 3p+\gamma\rightarrow 2s+2\gamma\rightarrow 1s+4\gamma$ transition. In this case we fix $E_{n_3}=E_{3p}$ and $E_{n_2}=E_{2s}$, then the resonant frequencies are $\omega=E_{4s}-E_{3p}$, $\omega'\equiv E_{3p}-E_{2s}$. Therefore, in the dipole approximation,
\begin{eqnarray}
\label{5.four}
U_{1s\, 4s}^{(4)}=\sum\limits_{n_1}\frac{\langle 1s|\vec{e'''}\vec{p}|n_1\rangle \langle n_1|\vec{e''}\vec{p}|2s\rangle \langle 2s|\vec{e'}\vec{p}|3p\rangle \langle 3p|\vec{e}\vec{p}|4s\rangle}{(E_{n_1}-E_{2s}+\omega'')(E_{2s}-E_{3p}+\omega')(E_{3p}-E_{4s}+\omega)}+\rm{other\,\, 23\,\, terms}\,
\end{eqnarray}
where 'other 23 terms' differ from the first one by permutation of matrix elements.
We again use the expressions for the one-photon probabilities:
\begin{eqnarray}
\label{6.four}
W_{4s-3p}^{(1\gamma)}=\frac{e^2\omega}{2\pi}\sum\limits_{\vec{e}}\int d\vec{\nu}\left|\langle 3p|\vec{e}\vec{p}|4s\rangle\right|^2
\\
W_{3p-2s}^{(1\gamma)}=\frac{e^2\omega'}{2\pi}\sum\limits_{\vec{e'}}\int d\vec{\nu'}\left|\langle 2s|\vec{e'}\vec{p}|3p\rangle\right|^2
\end{eqnarray}
and the equality
\begin{eqnarray}
\label{7.four}
\int\limits_0^{\omega_0}\frac{d\omega}{(\omega_0+\omega)^2+\frac{1}{4}\Gamma^2}=\frac{2\arctan{\frac{2\omega_0}{\Gamma}}}{\Gamma}=\frac{\pi}{\Gamma}
\end{eqnarray}
which is valid for the small (compared to $\omega_0$) values of $\Gamma$.

Then, we can get for the four-photon transition probability an expression:
\begin{eqnarray}
\label{8.four}
W_{4s-1s}^{(4\gamma)}=\frac{e^4}{4!(2\pi)^5}\sum\limits_{\vec{e''}\vec{e'''}}\int\limits_0^{\omega_{3p-2s}}\frac{W_{3p-2s}^{(1\gamma)}d\omega'}{(E_{2s}-E_{3p}+\omega')^2+\frac{1}{4}\Gamma_{4s}^2}\int\limits_0^{\omega_{4s-3p}}\frac{W_{4s-3p}^{(1\gamma)}d\omega}{(E_{3p}-E_{4s}+\omega)^2+\frac{1}{4}\Gamma_{3p}^2}\times
\nonumber
\\
\int d\vec{\nu''}\int d\vec{\nu'''}\int\omega''d\omega''\int\omega'''d\omega'''\left|\sum\limits_{n_1}\frac{\langle 1s|\vec{e'''}\vec{p}|n_1\rangle \langle n_1|\vec{e''}\vec{p}|2s\rangle}{E_{n_1}-E_{2s}+\omega''} +\rm{other\,\, 23\,\, terms} \right|^2\, ,
\end{eqnarray}
or, in equivalent form,
\begin{eqnarray}
\label{9.four}
W_{4s-1s}^{(4\gamma)}=\frac{e^4}{4!(2\pi)^5}\sum\limits_{\vec{e''}\vec{e'''}}\int\limits_0^{\omega_{3p-2s}}\frac{W_{3p-2s}^{(1\gamma)}d\omega'}{(E_{2s}-E_{3p}+\omega')^2+\frac{1}{4}\Gamma_{4s}^2}\int\limits_0^{\omega_{4s-3p}}\frac{W_{4s-3p}^{(1\gamma)}d\omega}{(E_{3p}-E_{4s}+\omega)^2+\frac{1}{4}\Gamma_{3p}^2}\times
\nonumber
\\
12^2\int d\vec{\nu''}\int d\vec{\nu'''}\int\omega''d\omega''\int\omega'''d\omega'''\left|\sum\limits_{n_1}\frac{\langle 1s|\vec{e'''}\vec{p}|n_1\rangle \langle n_1|\vec{e''}\vec{p}|2s\rangle}{E_{n_1}-E_{2s}+\omega''} + \sum\limits_{n_1}\frac{\langle 1s|\vec{e''}\vec{p}|n_1\rangle \langle n_1|\vec{e'''}\vec{p}|2s\rangle}{E_{n_1}-E_{2s}+\omega'''}\right|^2\, .
\end{eqnarray}
Finally, for the $4s\rightarrow 3p+\gamma\rightarrow 2s+2\gamma\rightarrow 1s+4\gamma$ emission process we get
\begin{eqnarray}
\label{10.four}
W_{4s-1s}^{(4\gamma)\rm{res1}}=\frac{3}{2}\frac{W_{4s-3p}^{(1\gamma)}}{\Gamma_{4s}}\frac{W_{3p-2s}^{(1\gamma)}}{\Gamma_{3p}}W_{2s-1s}^{(2\gamma)}\, .
\end{eqnarray}
Also the processes $4s\rightarrow 3s+2\gamma\rightarrow 2p+3\gamma\rightarrow 1s+4\gamma$ and $4s\rightarrow 3p+\gamma\rightarrow 2p+3\gamma\rightarrow 1s+4\gamma$ should be considered. Performing similar calculations we get
\begin{eqnarray}
\label{11.four}
W_{4s-1s}^{(4\gamma)\rm{res2}}=\frac{3}{2}\frac{W_{3s-2p}^{(1\gamma)}}{\Gamma_{3s}}\frac{W_{2p-1s}^{(1\gamma)}}{\Gamma_{2p}}W_{4s-3s}^{(2\gamma)} \equiv \frac{3}{2}\frac{W_{3s-2p}^{(1\gamma)}}{\Gamma_{3s}}W_{4s-3s}^{(2\gamma)}
\end{eqnarray}
and
\begin{eqnarray}
\label{12.four}
W_{4s-1s}^{(4\gamma)\rm{res3}}=\frac{3}{2}\frac{W_{4s-3p}^{(1\gamma)}}{\Gamma_{4s}}\frac{W_{2p-1s}^{(1\gamma)}}{\Gamma_{2p}}W_{3p-2p}^{(2\gamma)} \equiv \frac{3}{2}\frac{W_{4s-3p}^{(1\gamma)}}{\Gamma_{4s}}W_{3p-2p}^{(2\gamma)}\, .
\end{eqnarray}

Finally, for the four-photon decay of the $4s$ hydrogenic state within the two-photon approximation we recieve
\begin{eqnarray}
\label{13.four}
W_{4s-1s}^{\rm{total}}=W_{4s-1s}^{(2\gamma)}+\frac{3}{2}\frac{W_{3s-2p}^{(1\gamma)}}{\Gamma_{3s}}W_{4s-3s}^{(2\gamma)}+\frac{3}{2}\frac{W_{4s-3p}^{(1\gamma)}}{\Gamma_{4s}}W_{3p-2p}^{(2\gamma)}+\frac{3}{2}\frac{W_{4s-3p}^{(1\gamma)}}{\Gamma_{4s}}\frac{W_{3p-2s}^{(1\gamma)}}{\Gamma_{3p}}W_{2s-1s}^{(2\gamma)}\, .
\end{eqnarray}
where $W_{4s-1s}^{(2\gamma)}$ is the "pure" two-photon contribution to the $4s\rightarrow 1s$ decay (see Fig. 12). We remind that, in principle, this contribution is inseparable from the cascade contribution in Fig. 12. The numerical values for the cascade contributions with the two-photon links (three last terms in the right-hand side of Eq. (69)) are:
\begin{eqnarray}
\frac{3}{2}\frac{W_{3s-2p}^{(1\gamma)}}{\Gamma_{3s}}W_{4s-3s}^{(2\gamma)}\approx 0.00438791\, s^{-1}
\nonumber
\\
\frac{3}{2}\frac{W_{4s-3p}^{(1\gamma)}}{\Gamma_{4s}}W_{3p-2p}^{(2\gamma)}= 0.0284699\, s^{-1}
\\
\nonumber
\frac{3}{2}\frac{W_{4s-3p}^{(1\gamma)}}{\Gamma_{4s}}\frac{W_{3p-2s}^{(1\gamma)}}{\Gamma_{3p}}W_{2s-1s}^{(2\gamma)}= 0.611441\, s^{-1}
%\\
%\nonumber
%\Gamma_{4s}=W_{4s-3p}^{(1\gamma)}+W_{4s-2p}^{(1\gamma)}=0.043\cdot 10^8\, s^{-1}
\end{eqnarray}
These values are again not neglidgible compared to the contribution $W_{4s-1s}^{(2\gamma)}$. The latter can be estimated as $\sim 12\, s^{-1}$ \cite{J.Chluba}.

For the evaluation of the $4s\rightarrow 3s+2\gamma$ transition probability it is enough to consider the standard expression for the distribution function $dW(\omega)$ (Eqs (\ref{23})-(\ref{26})) with the replacement of the wave functions $2s\rightarrow 4s$, $1s\rightarrow 3s$. Note, that there are no intermediate $p$-states between $4s$ and $3s$ levels. The existence of the degenerate $4s$, $4p$ and $3s$, $3p$ levels does not change the result. The result of the calculations in the nonrelativistic limit is
\begin{eqnarray}
W_{4s-3s}^{(2\gamma)}=0.00879957\alpha^2(\alpha Z)^6\, r.u.=0.00292527\, s^{-1}.
\end{eqnarray}
The relativistic value is equal $0.002924794\,s^{-1} $. The relative difference of relativistic and nonrelativistic values is about $0.016 \%$. 

Similarly the nonrelativistic result for the $3p-2p$ two-photon transition rate can be obtained:
\begin{eqnarray}
W_{3p-2p}^{(2\gamma)} = 0.13618 \alpha^2(\alpha Z)^6\,a.u. = 0.045271 s^{-1}\,.
\end{eqnarray}
For the summation over the nonrelativistic hydrogen spectrum in the expressions for the two-photon transition rates we used the Green function method (see, for example, \cite{RZM}, \cite{LS} and also \cite{LSPS}). Here we give also the numerical result $W_{3s-2s}^{(2\gamma)}=0.194113\alpha^2(\alpha Z)^6\, r.u. =0.0645296\, s^{-1}$ with relative difference with the relativistic one $ 0.0082\%$, though this value does not enter directly to the $3p$, $4s$ decays.

\section{Conclusions}
In our paper we have considered the processes of multiphoton transitions for hydrogenic atom. Recent astrophysical investigations necessitate the detailed analysis of the multiphoton emission processes and, namely, of the separation of the "pure" multiphoton radiation and cascade emission. The "pure" two-photon emission leads to the photon escape from the matter and, thus, presents the formation mechanism for the background radiation.

We began with the standard QED derivation of the Lorentz profile for emission process. After this we have investigated the decay of the $3s$ level and showed the ambiguity of separation of the "pure" two-photon and cascade contributions. We demonstrated that the strict separation of the "pure" two-photon and cascade contributions for $3s$-level decay in hydrogen is impossible. Moreover, we show that even the approximate separation of these two decay channels cannot be achieved with an accuracy, required in modern astrophysical investigations (i.e. at 1\% level) of the recombination history of hydrogen in the early Universe. 

We formulated the "two-photon" approximation which make it possible to separate out cascade emission with two-photon links and show that this type of cascades gives the contribution to the two-photon transitions, i.e. to the radiation escape, comparable with the contribution of the "direct" two-photon transitions. On a basis of this approximation the decay of $3p$ level was described. It was shown that the cascade two-photon decay rates are comparable with the "pure" two-photon contribution of the $3s\rightarrow 1s+2\gamma(E1)$ channel. Unlike the $3s\rightarrow 1s+2\gamma(E1)$ case, where the contribution of the "pure" two-photon decay is nonseparable from the cascade contribution, for $3p\rightarrow 1s+3\gamma(E1)$ decay only the cascade contributions with the two-photon links should be taken into account. The reason is that the "pure" 3-photon contribution is beyond the accuracy of the "two-photon" approximation, adopted in this paper. The four-photon emission process of the $4s$ level was considered also. The result is: the decay rates for the cascades with the two-photon links are comparable with the contribution of the "direct" two-photon transitions. 

The main goal of our paper is the formulation of the "two-photon" approximation which allows for the rigarous incorporation of all types of the two-photon processes. This may be important for the more accurate astrophysical investigations of the cosmical radiation background.

\begin{center}
Acknowledgments
\end{center}
The authors acknowledge financial support from RFBR (grant Nr. 08-02-00026). The work of D.S. was supported by
the Non-profit Foundation ``Dynasty'' (Moscow). The authors acknowledge also 
the support by the Program of development of scientific potential of High School,
Ministry of Education and Science of Russian Federation (grant Nr. 2.1.1/1136 and goskontrakt $\Pi$1334).

\newpage
\begin{figure}[htp]
    \includegraphics[scale=0.55]{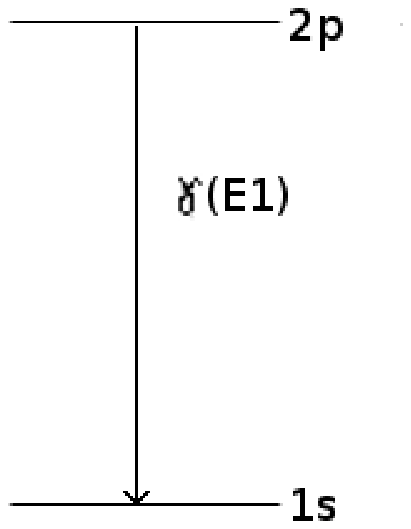}
    \\
{\citation
\\ Fig. 1. Schematic picture for the transition $2p\rightarrow 1s$ in hydrogen atom. Horizontal lines denote the electron levels, the vertical line with an arrow denotes the photon transition. Transition rate (in r.u.) is $W_{E1}^{(1\gamma)}(2p-1s)=732.722 m\alpha(\alpha Z)^4$. For hydrogen $Z=1$. Here $1\gamma$ denotes the number of photons and $E1$ describes the type of the photon.}
\end{figure}
\begin{figure}[htp]
%    \subfigure
    \includegraphics[scale=0.55]{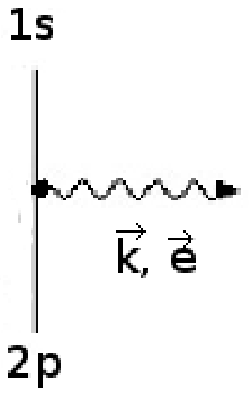}
    \\
{\citation
\\ Fig. 2. The Feynman graph, corresponding to the decay process Fig. 1. The solid line denotes the electron, the upper and lower parts of this line correspond to the final and initial electron states. The wave line with an arrow at the end denotes the emitted photon with momentum $\vec{k}$, frequency $\omega=|\vec{k}|$ and the polarization $\vec{e}$.}
\end{figure}
\begin{figure}[htp]
%    \subfigure
    \includegraphics[scale=0.6]{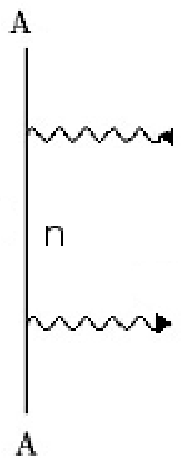}
    \\
{\citation
\\ Fig. 3. The Feynman graph describing the elactic photon scattering on an atomic electron. The indices $A$. $n$ correspond to the initial (final) and intermediate electron states. Notations are the same sa in Fig. 2.}
\end{figure}
\begin{figure}[htp]
%    \subfigure
\includegraphics[scale=0.4]{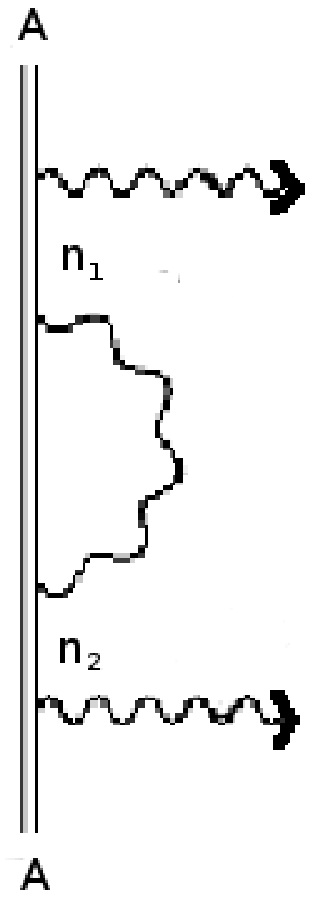}
    \\
{\citation
\\ Fig. 4. The Feynman graph corresponding to the lowest-order electron self-energy insertion into the electron propagator in Fig. 3. The internal wavy line denotes the photon propagator in the Feynman gauge.}
\end{figure}
\begin{figure}[htp]
%    \subfigure
\includegraphics[scale=0.6]{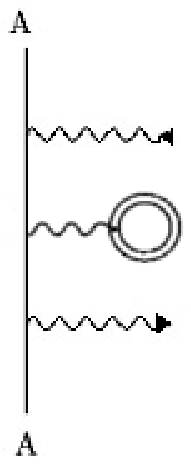}
    \\
{\citation
\\ Fig. 5. The vacuum polarization insertion in the electron propagator in Fig. 3. The notations are the same as in Figs. 3 and 4.}
\end{figure}
\begin{figure}[htp]
%    \subfigure
\includegraphics[scale=0.6]{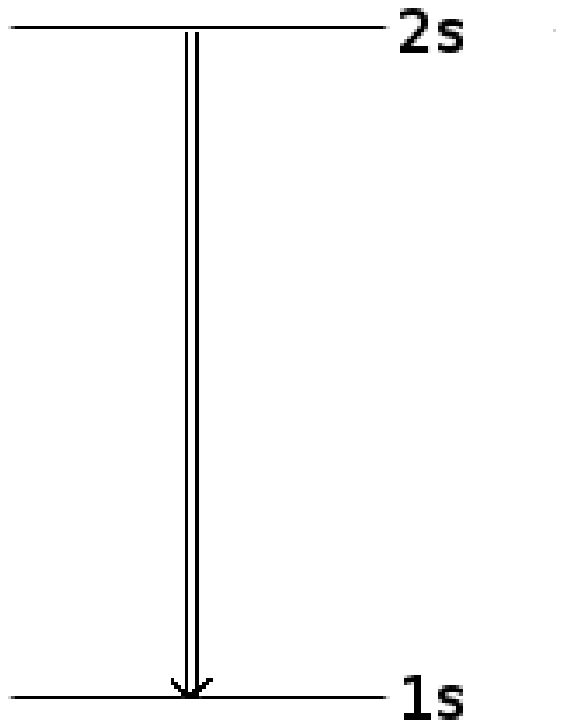}
    \\
{\citation
\\ Fig. 6. Schematic picture for the transition $2s\rightarrow 1s+2\gamma (E1)$. Double verticale line with arrow denotes the total two-photon transition (the "pure" two-photon) in rhis case.}
\end{figure}
\begin{figure}[htp]
%    \subfigure
\includegraphics[scale=0.6]{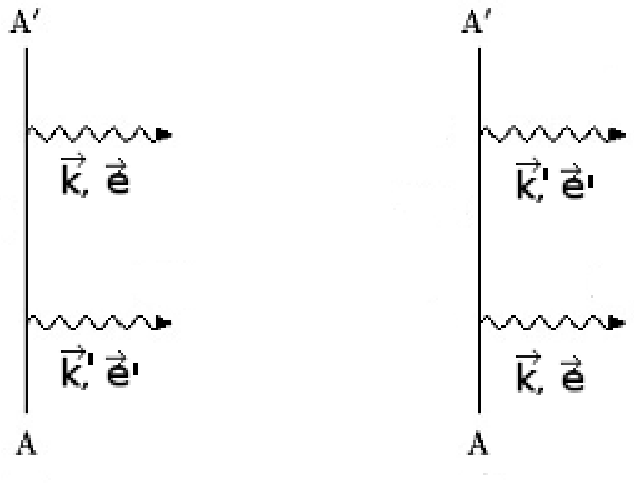}
    \\
{\citation
\\ Fig. 7. The Feynman graph for the two-photon emission process corresponding to the $A\rightarrow A'+2\gamma$ transition. All the notations are the same as in Fig. 2. Two graphs occur due to the permutation symmetry of the emitted photons.}
\end{figure}
\begin{figure}[htp]
%    \subfigure
\includegraphics[scale=0.6]{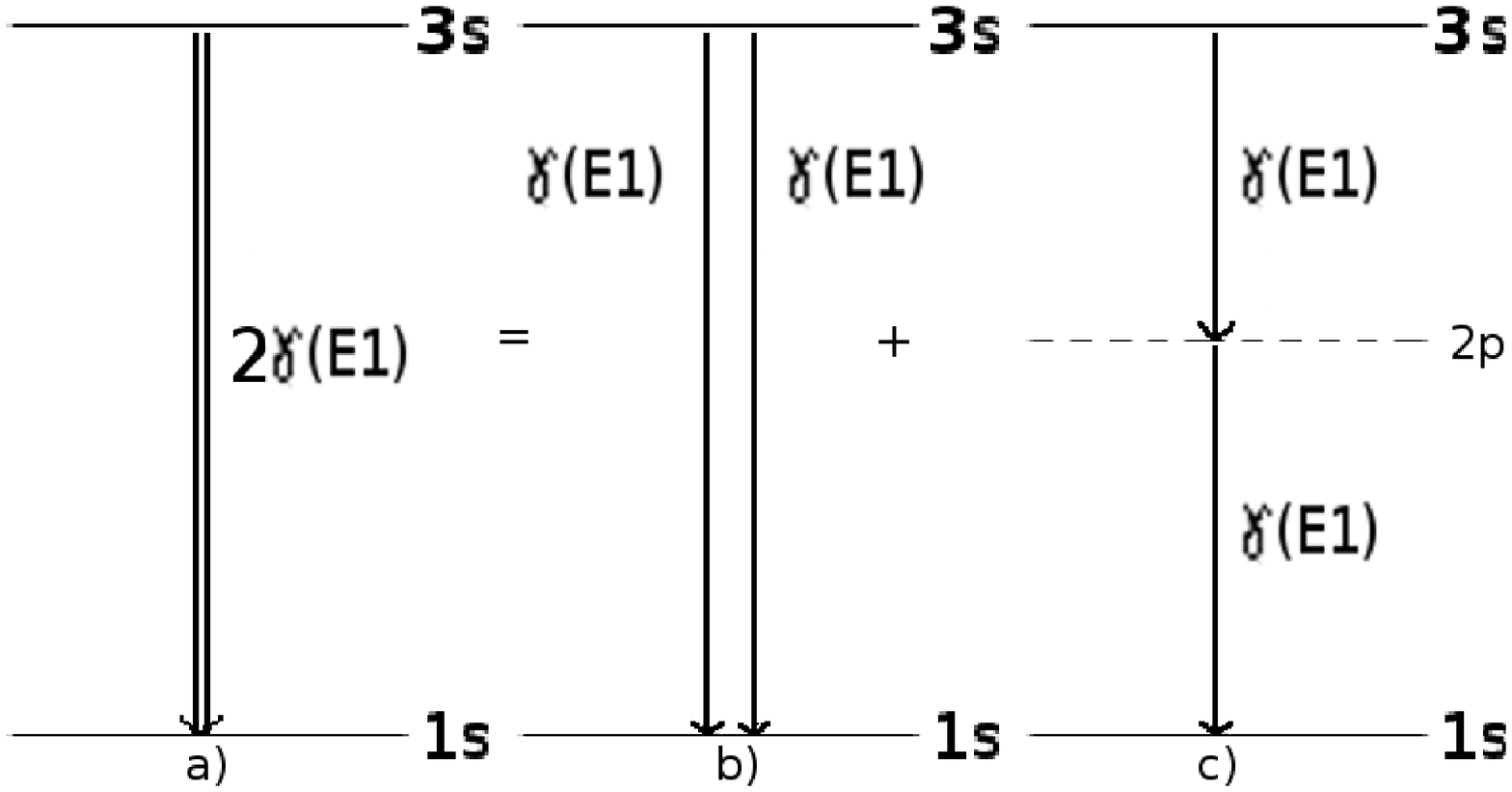}
    \\
{\citation
\\ Fig. 8. Schematic picture for the transition $3s\rightarrow 1s+2\gamma (E1)$. Double verticale line with arrow in Fig. 8 a) denotes the total two-photon transition, the two ordinary vertical lines in Fig. 8 b) sorrespond to the "pure" two-photon transition and the ordinary vertical lines with arrows in Fig. 8 c) describe the cascade photons. The horizontal dashed line in Fig. 8 c) denotes the intermediate energy level. One has to remember that Fig. 8 presents the decomposition of the amplitude, so the interference terms between channels 8 b) and 8 c) arise in the probability expression. Moreover, as it was explained in the text, actually the contributions 8 b) and 8 c) are inseparable.}
\end{figure}
\begin{figure}[htp]
%    \subfigure
\includegraphics[scale=0.6]{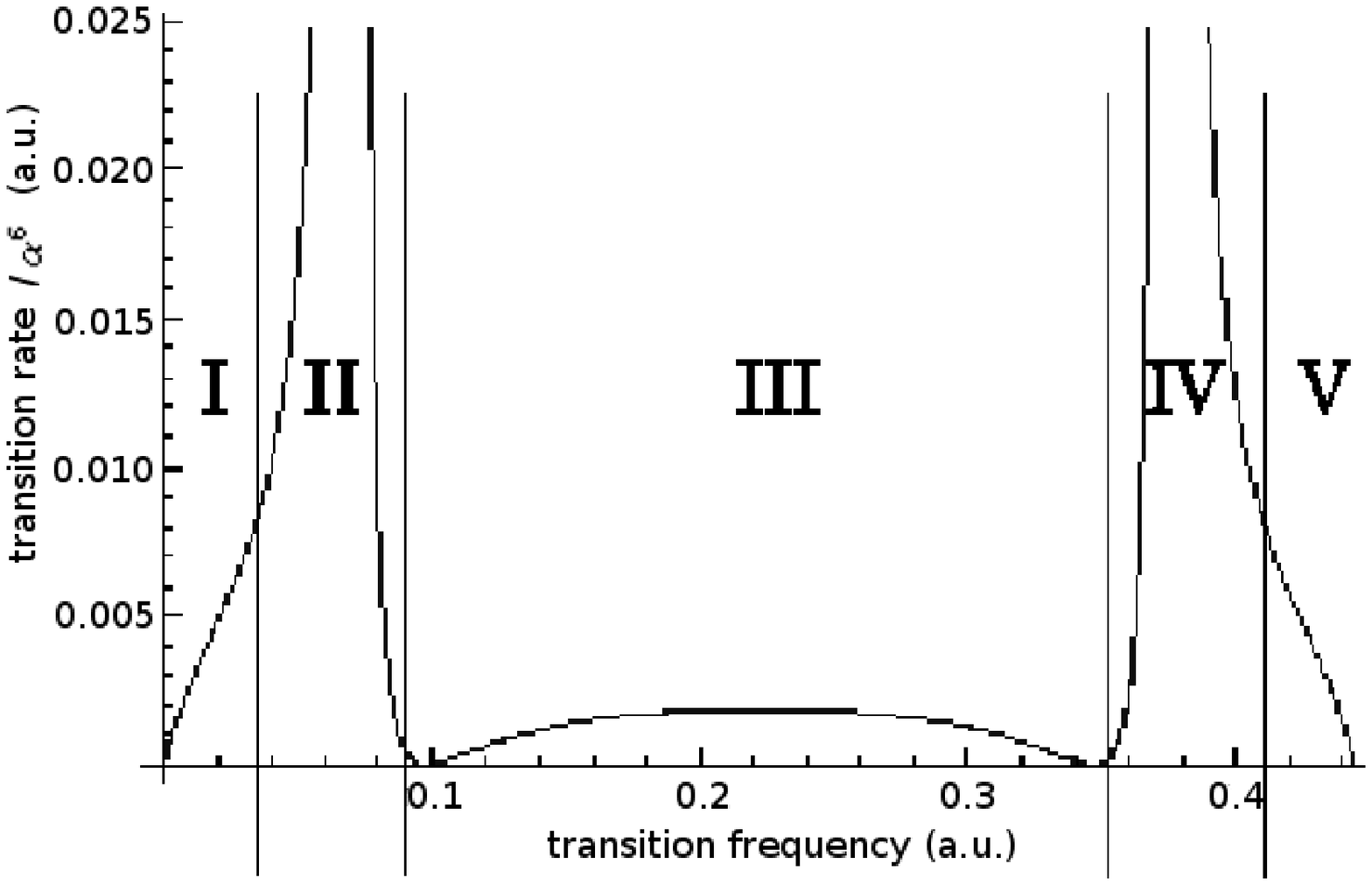}
    \\
    {\citation
\\ Fig. 9. The frequency distribution  $dW^{(2\gamma)}_{3s;1s}/d\omega$ for the total two-photon transition $3s\rightarrow 1s+2\gamma$ including cascade and "pure" two-photon transitions as functions of the frequency (in a.u.). The values $dW^{(2\gamma)}_{3s;1s}/d\omega$ divided by $\alpha^6$ ($\alpha$ is the fine structure constant) is plotted versus the frequency within the interval $[0,\omega_0]$. The boundaries for the frequency intervals {\bf I}-{\bf V} are also indicated as vertical lines.}
\end{figure}
\begin{figure}[htp]
%    \subfigure
\includegraphics[scale=0.4]{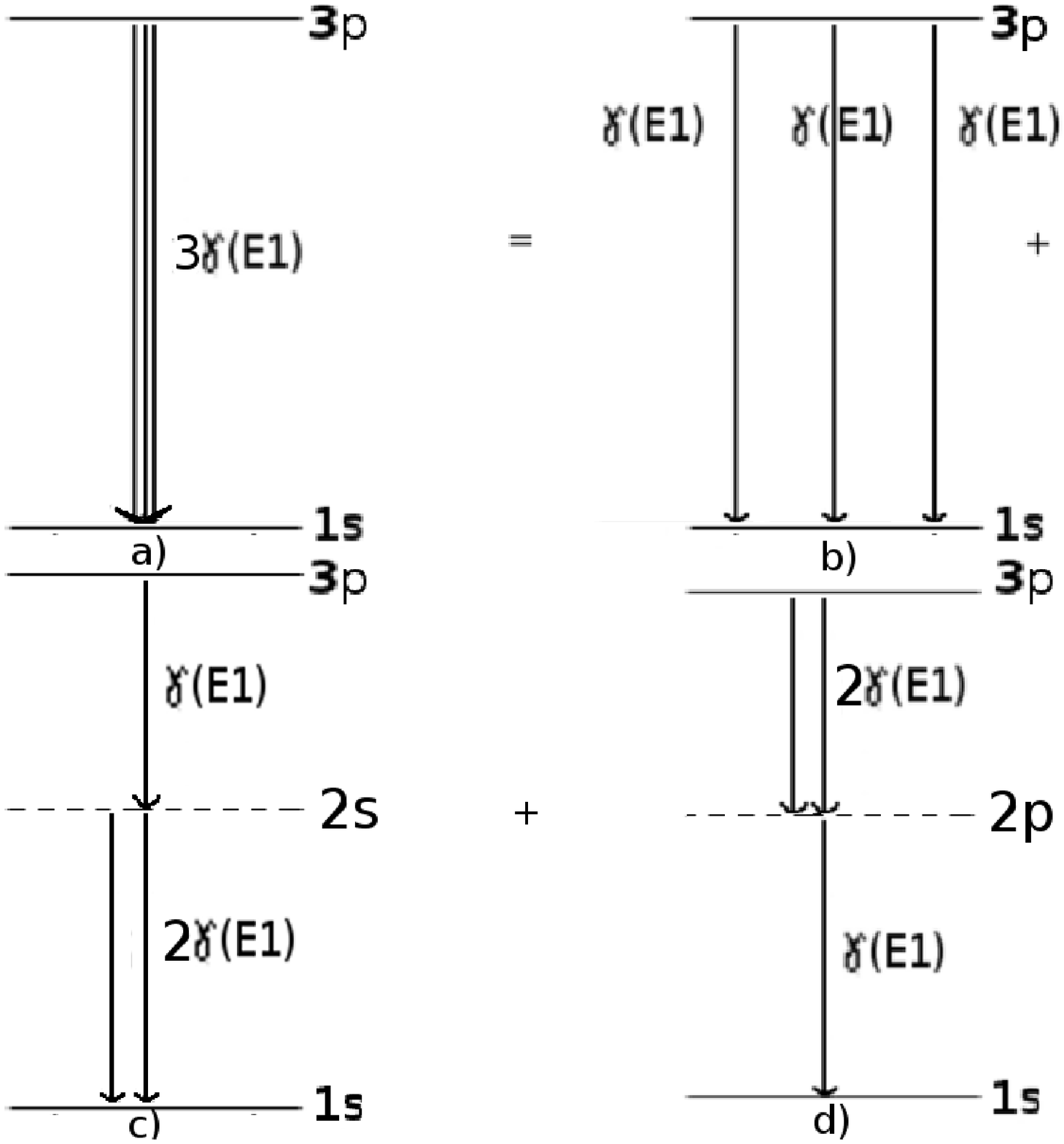}
    \\
    {\citation
\\ Fig. 10. Schematic picture for the transition $3p\rightarrow 1s+3\gamma (E1)$. Triple vertical line with arrow in Fig. 10 a) denotes the total three-photon contribution, the other notations are the same as in Fig. 6. In the "two-photon" approximation only the channels Fig. 10 c) and Fig. 10 d) contribute to the decay probability at the adopted level of accuracy.}
\end{figure}
\begin{figure}[htp]
%    \subfigure
\includegraphics[scale=0.8]{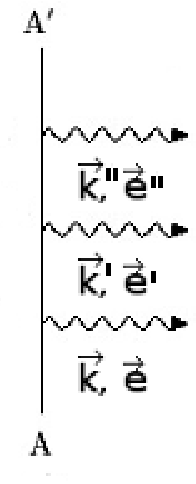}
    \\
{\citation
\\ Fig. 11. The Feynman graph for the three-photon emission process corresponding to the $A\rightarrow A'+3\gamma$ transition. All the notations are the same as in Fig. 2. The graphs with all permutations of the photon lines should be added.}
\end{figure}
\begin{figure}[htp]
%    \subfigure
\includegraphics[scale=0.6]{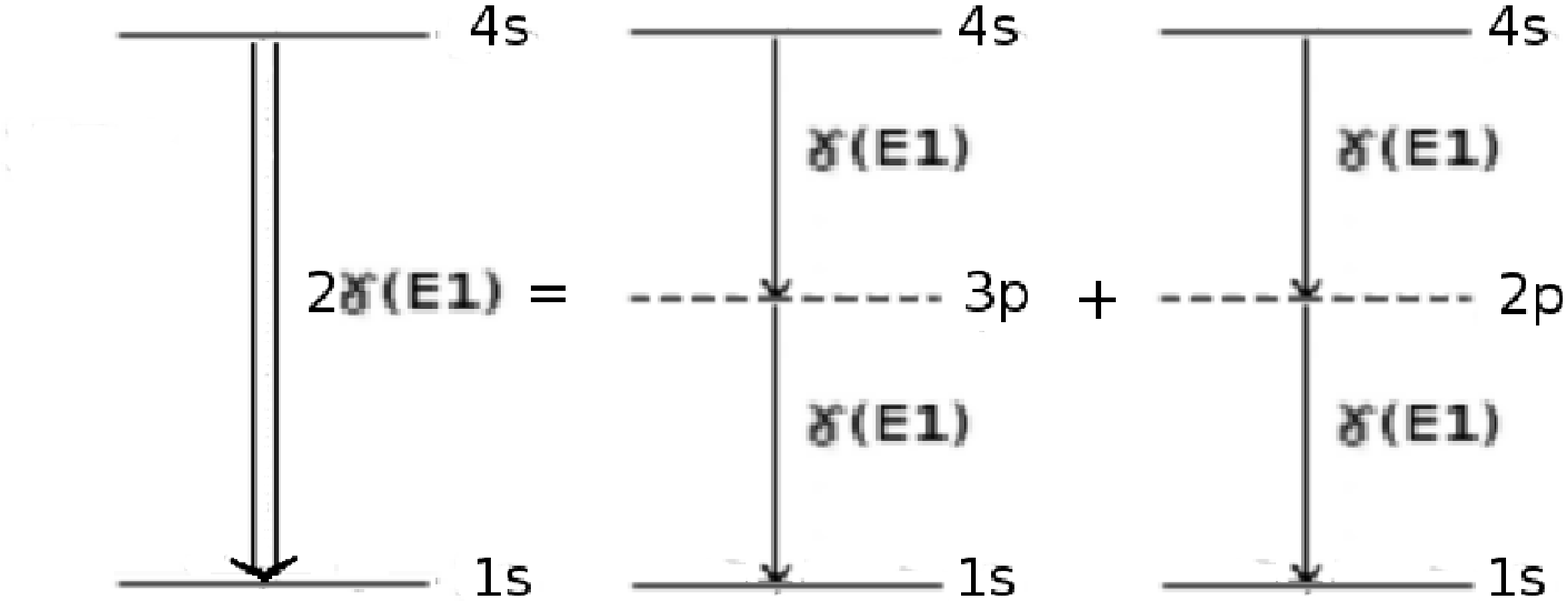}
    \\
{\citation
\\ Fig. 12. Schematic picture of the two-photon decay of the state $4s$. Notations are the same as in Fig. 7}
\end{figure}
\begin{figure}[htp]
%    \subfigure
\includegraphics[scale=0.6]{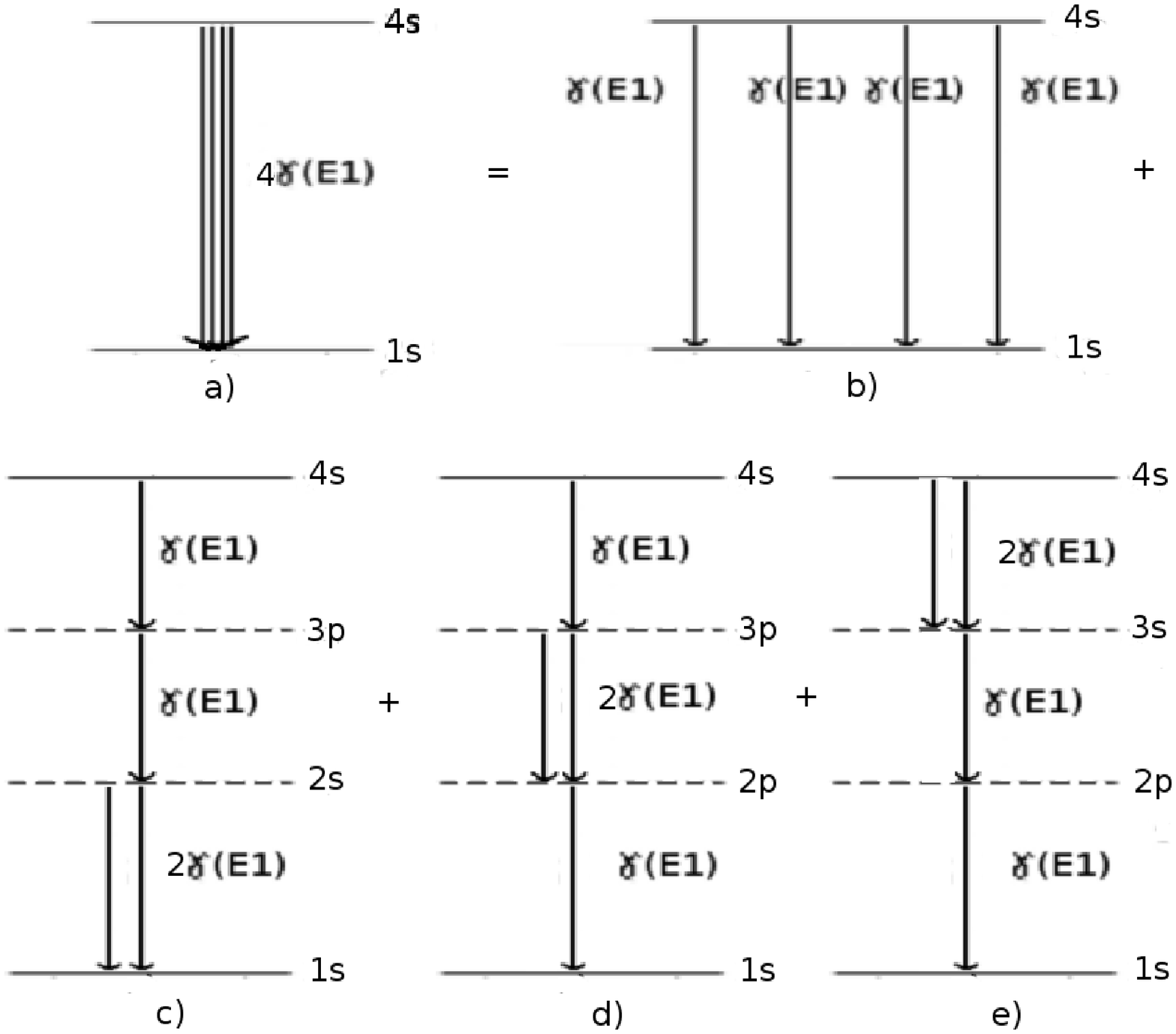}
    \\
{\citation
\\ Fig. 13. Schematic picture of the four-photon decay of the state $4s$. Quadruple vertical line denotes the total 4-photon contribution. In the two-photon approximation only c), d) and e) schemes contribute.}
\end{figure}
\end{document}